\documentclass[english, apj]{emulateapj}
\usepackage[T1]{fontenc}
\usepackage[latin9]{inputenc}
\usepackage{array}
\usepackage{rotating}
\usepackage{units}
\usepackage{textcomp}
\usepackage{amsmath}
\usepackage{amsbsy}
\usepackage{amstext}
\usepackage{graphicx}
\usepackage{url}
\usepackage{babel}
\usepackage[backref,breaklinks,colorlinks,citecolor=blue]{hyperref}
\usepackage[all]{hypcap}

\makeatletter

\newcommand{\msun}{\mbox{$\,{\rm M}_\odot$}}

\begin{document}

\title{Globular Cluster Streams as Galactic High-Precision Scales -- The Poster Child Palomar\,5}

\author{Andreas H.W. K\"upper\altaffilmark{1,2}, Eduardo Balbinot\altaffilmark{3}, Ana Bonaca\altaffilmark{4}, Kathryn V. Johnston\altaffilmark{1}, David W. Hogg\altaffilmark{5}, Pavel Kroupa\altaffilmark{6} Basilio X. Santiago\altaffilmark{7,8}}
\altaffiltext{1}{Department of Astronomy, Columbia University, 550 West 120th Street, New York, NY 10027, USA}
\altaffiltext{2}{Hubble Fellow}
\altaffiltext{3}{Department of Physics, University of Surrey, Guildford GU2 7XH, UK}
\altaffiltext{4}{Department of Astronomy, Yale University, New Haven, CT 06511, USA}
\altaffiltext{5}{Center for Cosmology and Particle Physics, Department of Physics, New York University, 4 Washington Place,
New York, NY 10003, USA}
\altaffiltext{6}{Helmholtz-Institut f\"ur Strahlen- und Kernphysik (HISKP), University of Bonn, Nussallee 14-16, 53115 Bonn, Germany}
\altaffiltext{7}{Departamento de Astronomia, Universidade Federal do Rio Grande do Sul, Av. Bento Gon\c{c}alves 9500, Porto Alegre 91501-970, RS, Brasil}
\altaffiltext{8}{Laborat\'orio Interinstitucional de e-Astronomia - LIneA, Rua Gal. Jos\'e Cristino 77, Rio de Janeiro, RJ - 20921-400, Brasil}
\email{Correspondence to: akuepper@astro.columbia.edu}

\begin{abstract}
Using the example of the tidal stream of the Milky Way globular cluster Palomar\,5 (Pal\,5), we demonstrate how observational data on tidal streams can be efficiently reduced in dimensionality and modeled in a Bayesian framework. Our approach combines detection of stream overdensities by a Difference-of-Gaussians process with fast streakline models of globular cluster streams and a continuous likelihood function built from these models. Inference is performed with Markov chain Monte Carlo. By generating $\approx 10^7$ model streams, we show that the unique geometry of the Pal\,5 debris yields powerful constraints on the solar position and motion, the Milky Way and Pal\,5 itself. All 10 model parameters were allowed to vary over large ranges without additional prior information. Using only readily-available SDSS data and a few radial velocities from the literature, we find that the distance of the Sun from the Galactic Center is $8.30\pm0.25$\,kpc, and the transverse velocity is $253\pm16$\,km\,s$^{-1}$. Both estimates are in excellent agreement with independent measurements of these two quantities. Assuming a standard disk and bulge model, we determine the Galactic mass within Pal\,5's apogalactic radius of 19\,kpc to be $(2.1\pm0.4)\times10^{11}\msun$.
Moreover, we find the potential of the dark halo with a flattening of $q_z = 0.95^{+0.16}_{-0.12}$ to be essentially spherical -- at least within the radial range that is effectively probed by Pal\,5. We also determine Pal\,5's mass, distance and proper motion independently from other methods, which enables us to perform vital cross-checks. Our inferred heliocentric distance of Pal\,5 is $23.6^{+0.8}_{-0.7}$\,kpc, in perfect agreement with, and more precise than estimates from isochrone fitting of deep HST imaging data. We conclude that finding and modeling more globular cluster streams is an efficient way for mapping out the structure of our Galaxy to high precision.  With more observational data and by using additional prior information, the precision of this mapping can be significantly increased.
\end{abstract}


\section{Introduction}

Every satellite orbiting a host galaxy produces tidal debris. Yet, we find debris only around very few globular clusters or dwarf galaxies. This is because the amount and shape of tidal debris differs substantially from satellite to satellite. That is, every satellite leaves its own unique trace in the gravitational potential of its host. This being said, we can expect tidal debris to contain substantial information about the satellite, its orbit, and its host. Using the example of the most prominent globular cluster stream discovered so far, Palomar\,5, we are going to demonstrate in the following how this information can be extracted through sophisticated stream detection, numerical modeling and Bayesian inference.\\\\
With the onset of the era of survey science, detections of tidal debris have become numerous -- as extended overdensities in surface density; first from photographic plates \citep{Grillmair95, Kharchenko97, Lehmann97, Leon00, Lee03}, and more recently in the data from the Sloan Digital Sky Survey \citep{Odenkirchen01, Odenkirchen03, Grillmair06a, Grillmair06b, Grillmair06c, Grillmair06d, Belokurov06b, Belokurov07}. This number grew as the homogeneity and coverage increased with subsequent data releases \citep{Grillmair09, Balbinot11, Bonaca12, Grillmair13, Bernard14, Grillmair14}. Newly available imaging surveys like ATLAS \citep{Koposov14} or PAndAS \citep{Martin14} add to this growing list of tidal features. Some tidal streams were also, or solely, found as coherent structures in velocity space \citep{Helmi99, Newberg09, Williams11, Carlin12, Sesar13, Sohn14}. With the richness of ongoing and future surveys like APOGEE, DES, LAMOST, PanSTARRS, RAVE, SkyMapper, Gaia, and LSST \citep{Perryman01, Steinmetz06, Keller07, Ivezic08, Kaiser10, Eisenstein11, Rossetto11, Cui12, Zhao12}, we will be supplied with ever growing amounts of observational data on tidal debris of known and unknown satellites within the Milky Way disk and in the halo.\\\\
Despite this amount of data -- available or yet to come -- few attempts have been made to exploit the wealth of information encoded in these coherent structures. Most investigations studied theoretical cases (e.g.~\citealt{Murali99, Eyre09a, Eyre09b, Varghese11, Penarrubia12, Sanders14}), and only a few have actually tried to model the observations directly (e.g.~\citealt{Dehnen04, Fellhauer07, Besla10, Koposov10, Lux13}), as appropriate modeling techniques are only becoming available now. The two most noteworthy examples for streams that have been modeled by utilizing the observational data self-consistently are the Sagittarius stream and the GD-1 stream, which will be discussed in Sec.~\ref{ssec:discussion} in detail. The modeling of these two tidal streams have demonstrated the power (but also the caveats) of using tidal streams to infer the mass and shape of the Milky Way's gravitational potential.\\\\
The reason for the relatively low number of fully modeled tidal streams is twofold: first of all, generating a realistic model of a satellite's debris usually requires $N$-body simulations of some sort, which are computationally expensive and therefore do not allow for large model-parameter scans. However, a realistic stream model is necessary, since it has been shown that streams do not strictly follow progenitor orbits \citep{Johnston98, Helmi99}, and that fitting a stream with a single orbit can lead to significant biases in potential estimates \citep{Lux13, Sanders13}. Angle-action approaches for generating stream realizations are a promising way of reducing the computational needs \citep{Sanders14, Bovy14}. Here we will make use of restricted three-body models, so-called \textit{streaklines}, that allow us to generate a realistic stream model at small computational cost \citep{Kupper12}. The second reason is that comparisons of observational data to theoretical models are not straightforward. This is in most cases due to the faintness of the structures, that is, due to the heavy foreground and background contaminations from the Galaxy itself. Probabilistic approaches to such a problem, however, require a manifold of stream models, which is why most stream modeling approaches so far are only applied to (uncontaminated and error-free)  simulations of satellite streams \citep{Lux13, Bonaca14, Deg14} or remain qualitative \citep{Belokurov06a, Fellhauer06, Lux12, Zhang12, Vera13}. Newly available and computationally inexpensive generative models finally allow us to make such probabilistic statements on the model parameters describing observed tidal streams.\\\\
The problems mentioned above can be alleviated, if both observational data and stream models are simplified in a sensible way, for example, when individual, definite members of a satellite's debris can be identified and when their phase-space coordinates can be measured with high precision (e.g., RR Lyraes or blue horizontal branch stars; \citealt{Vivas06, Abbas14, Belokurov14}). With the \texttt{REWINDER} method, \citet{Price13} demonstrated that, in this case, orbits of stream stars can be traced back in time to the point where they escaped from the progenitor. With only a few stream stars the authors demonstrated that they are able to tightly constrain their model parameters \citep{Price14}. However, such an identification of stream members may often not be possible due to the sparsity of tracers or due to the unavailability of accurate phase-space information.\\\\
In this work, we present an alternative way to address both of these problems, i.e.~that allows us to quickly generate thousands of stream models, and to compare them to well-established parts of an otherwise heavily contaminated stream in the halo of the Milky Way even if large parts of the phase-space information is missing. We will make use of a new method for detecting stream overdensities, computationally efficient \textit{streakline} models \citep{Kupper12}, and the \texttt{FAST-FORWARD} method, a Bayesian framework developed in \citet{Bonaca14} for constraining galactic potentials by forward modeling of tidal streams (see Sec.~\ref{ssec:streaklines}).\\\\
Here we are going to apply a variant of the \texttt{FAST-FORWARD} method to the tidal debris of the low-mass globular cluster Palomar\,5 (Pal\,5; see Sec.~\ref{ssec:observations} for a detailed description). Due to its unique position high above the Galactic bulge and due to its current orbital phase of being very close to its apogalacticon, the tidal tails of this cluster are so prominent in SDSS that Pal\,5 can be considered to be the poster child of all globular cluster streams. Its tails stand out so prominently from the foreground/background that substructure can be seen within the stream with high significance (see Sec.~\ref{ssec:overdensities}, and \citealt{Carlberg12b}). Using a \textit{Difference-of-Gaussians} process, we will detect overdensities, measure their positions and extents, and then use them in our modeling to compare our models to these firmly established parts of the Pal\,5 stream (Sec.~\ref{ssec:comparison}).\\\\
Understanding the origin of such stream overdensities may give us detailed insight to a satellite's past dynamical evolution. \citet{Combes99} found that tidal shocks from the Galactic disk or bulge shocks could introduce substructure in dynamically cold tidal tails. However, \citet{Dehnen04} used $N$-body simulations of Pal\,5 to show that disk or bulge shocks are not the origin of the observed overdensities. Instead, the authors suggested that larger perturbations such as giant molecular clouds or spiral arms may have caused this substructure. In subsequent publications, several other scenarios for substructure formation within tidal streams have been discussed. A promising source of frequent and strong tidal shocks should be given through dark matter subhalos orbiting in the Galactic halo \citep{Ibata02, Johnston02, Siegal08, Carlberg09}. Encounters with these subhalos should be the more frequent the lower the subhalo mass -- and the hope is that, with better observational data, tidal streams may be used as detectors for these dark structures \citep{Yoon11, Carlberg12a, Erkal15}. In a promising first attempt, \citet{Carlberg12b} quantified substructures in Pal\,5's tidal tails and compared the number of detected gaps in the stream with expectation of $N$-body simulations. The authors concluded that some structures in the Pal\,5 stream are likely to be of such an origin. A different approach was taken by \citet{Comparetta11}, who suggested that Jeans instabilities could form periodic patterns in cold tidal streams. However, \citet{Schneider11} showed that tidal streams are gravitationally stable, and hence that Pal\,5's overdensities cannot be of this origin. As an alternative to tidal shocks or gravitational instabilities, \citet{Kupper08a} and \citet{Just09} found that epicyclic motion of stars along tidal tails can create periodic density variations in the tails when they are generated assuming a continuos flow of stars through the Lagrange points (see also Sec.~\ref{sssec:escape}). \citet{Kupper10b} and \citet{Kupper12} showed that these regular patterns of overdensities and underdensities are also present in tidal tails of clusters on eccentric orbits about their host galaxy, and that they are most pronounced when the cluster is in the apocenter of its orbit. Moreover, \citet{Mastrobuono12} demonstrated with $N$-body simulations that clusters on Pal\,5-like orbits can produce regularly spaced overdensities in their tails. In the following, we are going to show with our modeling of the Pal\,5 stream that several of the observed overdensities closer to the cluster are very likely to be due to such epicyclic motion. These overdensities help us to find the correct model of Pal\,5 as they stick out more significantly from the foreground/background than the rest of the stream.\\\\
This paper is organized as follows: in the next Section we are going to introduce our methodology, that is, we will describe the observational data this work is based on (Sec.~\ref{ssec:observations}), and explain how we extract the locations of stream overdensities from this data (Sec.~\ref{ssec:overdensities}), as we are going to fit our models to these overdensities. The modeling procedure for tidal streams is presented in Sec.~\ref{ssec:streaklines}, where we describe how we generate \textit{streakline} models of Pal\,5's tidal tails, how we compare the models to the observational data using a Bayesian approach, and how we scan the parameter space for the most probable model parameters using the Markov chain Monte Carlo method. In Sec.~\ref{sec:results}, we show the results of this modeling, first for an analog stream generated from an $N$-body simulation (Sec.~\ref{ssec:resultsanalog}), and then for the real observations of Pal\,5 (Sec.~\ref{ssec:resultspal5}). The last part of this paper is a discussion of the most probable parameters resulting from our modeling, with a comparison to results from other studies (Sec.~\ref{ssec:discussion}), followed by summary and conclusions in Sec.~\ref{sec:conclusions}.

\section{Method}\label{sec:method}

\subsection{Pal\,5 observational data}\label{ssec:observations}

\begin{figure}
\centering
\includegraphics[width=0.45\textwidth]{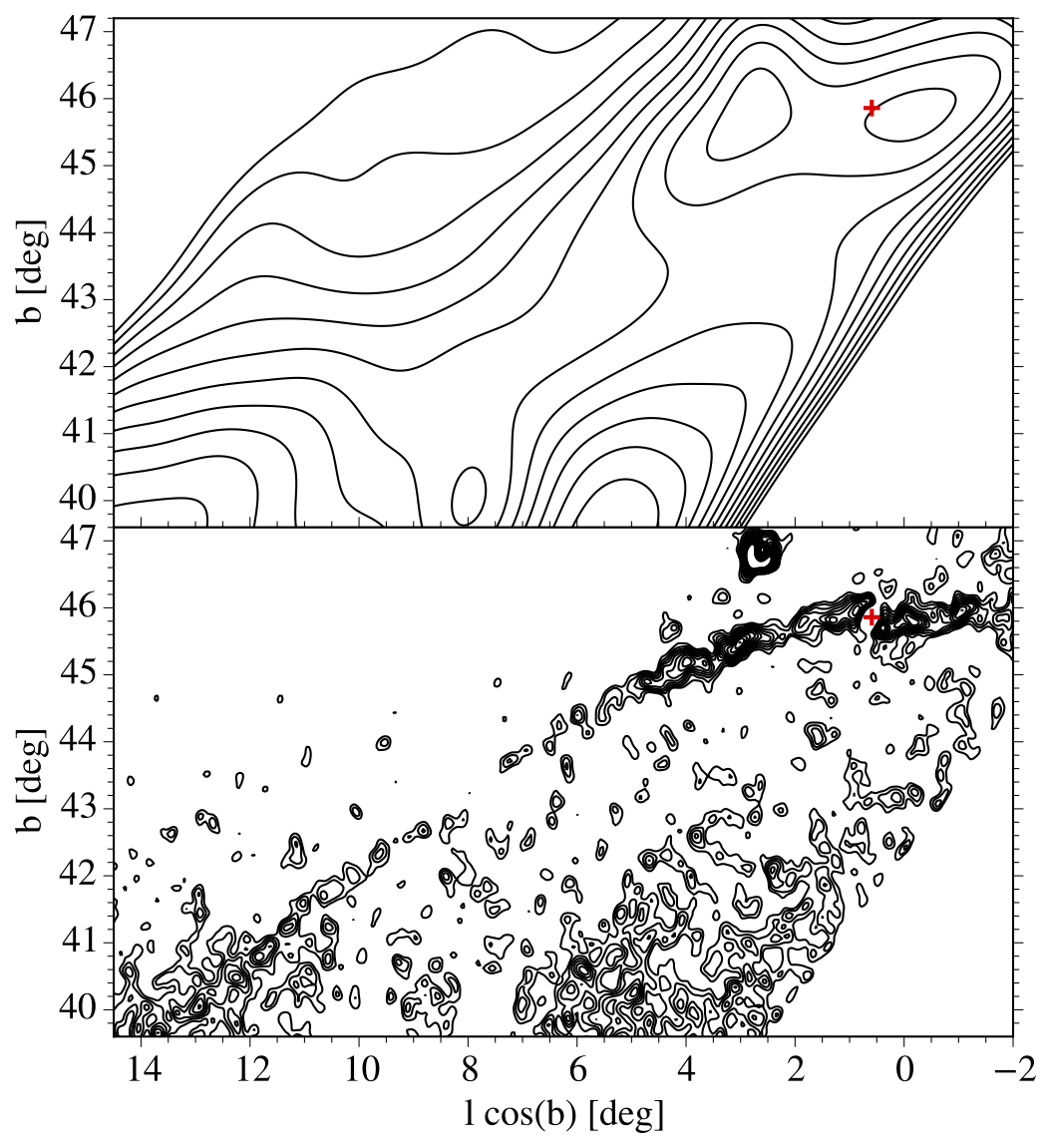}\\
\caption{Probability-density contours of Pal\,5 stars from our matched-filter analysis of the SDSS data set. Pal\,5 lies at (l\,$\cos\left(\mbox{b}\right)$, b) of (0.59, 45.86) deg and has been masked out (red cross). The leading tail is to the right of the cluster, the trailing to the left. The large structure at (2.64, 46.80) deg is the residual of the foreground cluster M\,5. Upper panel: density contours smoothed with a Gaussian kernel with a FWHM of 1.8\,deg showing the large scale variations across the stream. Lower panel: same as upper panel but for a FWHM of 0.23\,deg. Smaller kernel sizes reveal more substructure but also generate more statistical artifacts. For our analysis we subtracted the upper panel from the lower panel (see Fig.~\ref{od_real}).}
\label{SDSS}
\end{figure}

Being one of the faintest and most extended globular clusters of the Milky Way, Palomar 5 was first discovered on photographic plates from the Palomar Observatory Sky Survey \citep{Abell55}. Years later and to the great benefit of tidal stream research, it happened to lie within the footprint of the Sloan Digital Sky Survey (SDSS), although unfortunately quite close to the southern edge of the survey footprint. Using a matched filter on SDSS commissioning data, \citet{Odenkirchen01} detected the tidal tails of Pal\,5 stretching 2.6 deg on the sky. This marked the discovery of the first coherently mapped tidal stream. With subsequent data releases, the tails were found to span 22 deg within the SDSS footprint \citep{Rockosi02, Odenkirchen03, Grillmair06d}, where the trailing tail can be traced for about 18.5 deg, while the leading tail is cut off by the edge of the survey at a length of only 3.5 deg. Successive attempts to detect the trailing tail beyond this extent, e.g., by using neural networks \citep{Zou09}, had not been successful. However, with SDSS Data Release 8 the photometric uniformity and coverage of the Pal\,5 stream increased  \citep{Aihara11}. In this data, \citet{Carlberg12b} found the trailing stream to extend out to 23.2 deg from the cluster center, where the stream runs off the footprint.\\\\
To produce a density map of the Pal\,5 stream for our analysis, we apply the method outlined in \citet{Balbinot11} to the SDSS DR9 data \citep{Ahn12}. We select objects brighter than $g=22.5$ and classified as stars by the SDSS pipeline. Additional color cuts were performed following the same recipe as in \citet{Odenkirchen03}. The resulting map is represented on a grid of $0.03\times0.03$\,deg bins. In Fig.~\ref{SDSS} we show this map smoothed with a Gaussian kernel, using two different kernel widths (0.23 and 1.8\,deg, respectively). In the lower panel, Pal\,5's tidal tails can be clearly seen as they emanate from the cluster at (l\,$\cos\left(\mbox{b}\right)$, b) = (1, 46)\,deg and extend to about (-3, 46)\,deg for the leading tail, and about (15, 39)\,deg for the trailing tail. Both tails are limited by the SDSS footprint, but the trailing tail gets significantly fainter at the far end due to a wider dispersion of the stream, and also due to the larger extinction and contamination from the bulge at lower Galactic latitudes.\\\\
As can be seen in the lower panel of Fig.~\ref{SDSS}, the foreground/background around the stream shows many density enhancements of Pal\,5-like stars. The most obvious of these features is the residual of the foreground cluster M\,5 at (3, 47)\,deg. In order to distinguish between parts that belong to the Pal\,5 stream and contaminations, we are going to extract regions within the SDSS footprint in which Pal\,5-like stars are statistically over-dense compared to foreground/background fluctuations.

\subsubsection{Overdensity extraction}\label{ssec:overdensities}

\begin{figure}
\centering
\includegraphics[width=0.45\textwidth]{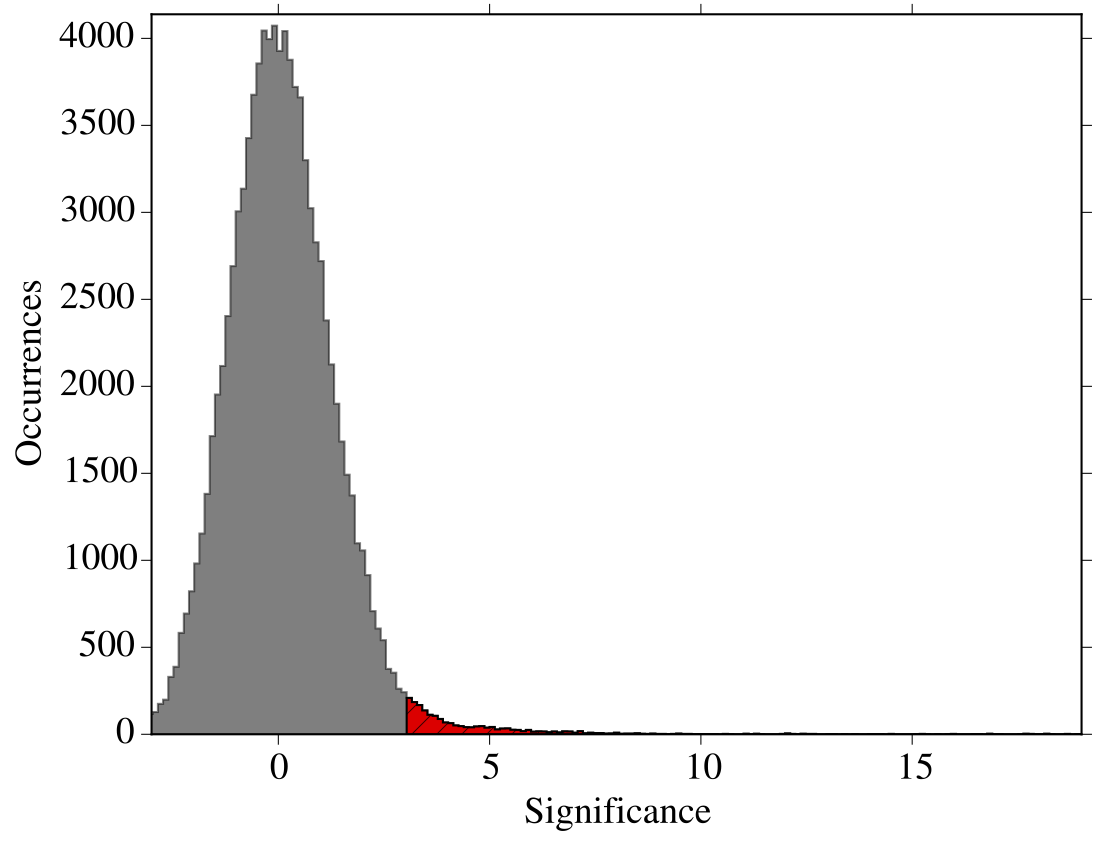}
\caption{Histogram of significances computed for each pixel in our residual map using Eq.~\ref{eq:significance}. The distribution of significances is approximately normal-distributed. Overdensities in the residual map are detected by searching for regions with more than 10 connected pixels with significance $>3\sigma$.}
\label{significances}
\end{figure}

\begin{figure*}
\centering
\includegraphics[width=0.9\textwidth]{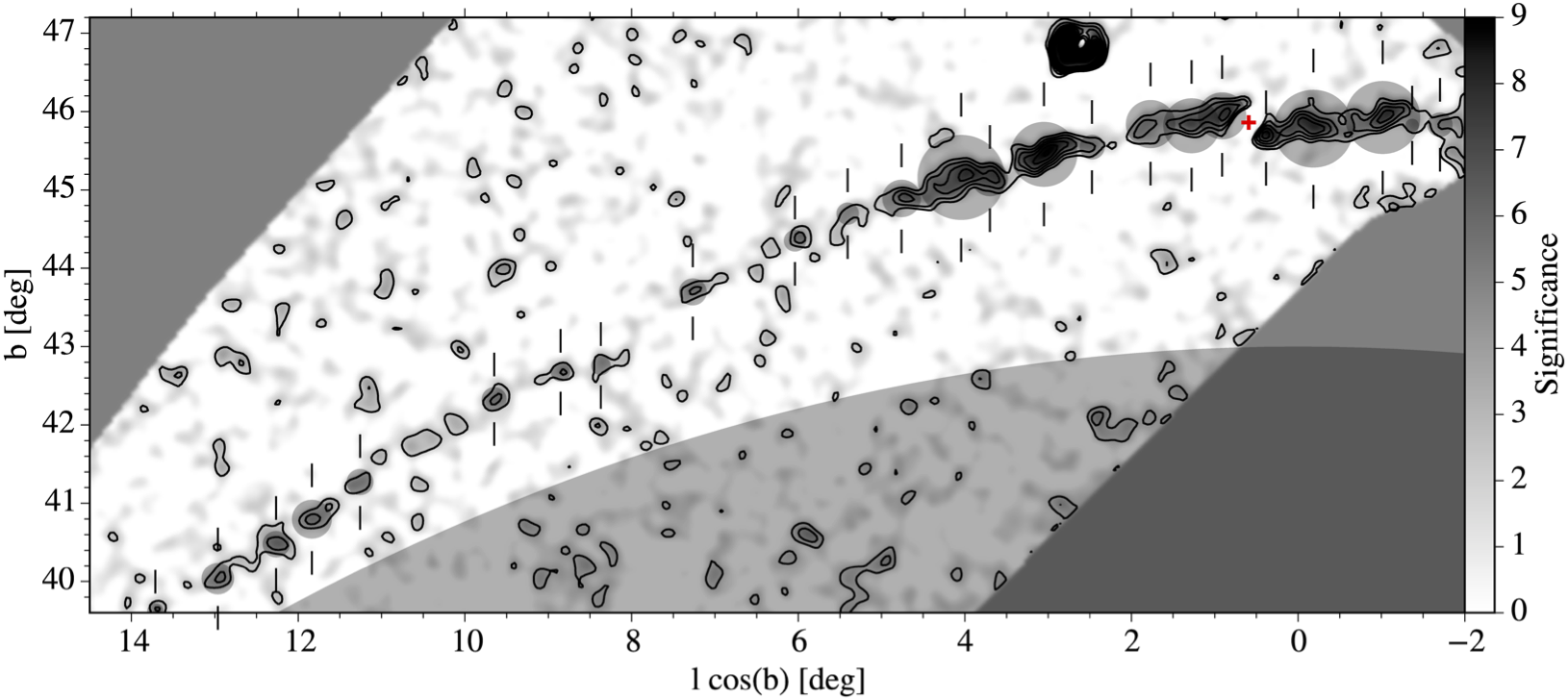}
\caption{Residual map of the region around Pal\,5 (red cross) as obtained by subtracting the upper panel of Fig.~\ref{SDSS} from the lower panel. The greyscale shows the local significance of each pixel as given from Eq.~\ref{eq:significance}. Barycenters of the 24 detected overdensities are marked by pairs of vertical lines, and their FWHM are illustrated by disks (see also Tab.~\ref{tab:overdensities}). We excluded the grey shaded region from our analysis, where the contaminations from the Galactic bulge, or border effects produce random features. Contours correspond to significances of [2, 3.5, 5, ...].
}
\label{od_real}
\end{figure*}

\begin{table}
 \centering
 \label{tab:overdensities}
 \caption{Overdensities}
 \begin{tabular}{ccccc}
 Name & l\,$\cos(\mbox{b})$ [deg] & b [deg] & Significance & FWHM [deg]\\
 \hline
T19 &14.66 &  39.19 &  3.48 &  0.15\\
T18 &13.71 &  39.64 &  3.41 &  0.14\\
T17 &12.96 &  40.06 &  3.96 &  0.28\\
T16 & 12.28 &  40.48 &  4.38 &  0.23\\
T15 &11.77 &  40.84 &  4.36 &  0.34\\
T14 &11.28 &  41.25 &  4.03 &  0.23\\
T13 & 9.65 &  42.33 &  4.34 &  0.21\\
T12 & 8.85 &  42.68 &  3.61 &  0.16\\
T11 & 8.39 &  42.75 &  3.48 &  0.17\\
T10 &7.26 &  43.69 &  4.23 &  0.24\\
T9 & 6.02  &   44.37 &  3.93 &  0.19\\
T8 & 5.43  &   44.68 &  3.76 &  0.19\\
T7 & 4.74  &   44.89 &  5.29 &  0.33\\
T6 & 4.14  &   45.11 &  7.43 &  0.74\\
T5 & 3.70  &   45.15 &  5.22 &  0.31\\
T4 & 3.04  &   45.45 &  10.31 &  0.57\\
T3 & 2.50  &   45.54 &  4.17 &  0.22\\
T2 & 1.82  &   45.80 &  5.06 &  0.42\\
T1 & 1.27  &   45.83 &  6.67 &  0.48\\
\hline
L1 & -0.06 &   45.79 &  7.94 &  0.68\\
L2 & -0.95 &   45.89 &  7.93 &  0.64\\
L3 &-1.37 &   45.82 &  3.29 &  0.11\\
L4 & -1.73 &   45.84 &  3.93 &  0.21\\
L5 & -2.05 &   46.13 &  3.16 &  0.11\\
 \end{tabular}
\end{table}
\citet{Carlberg12b} detected and characterized statistically significant density variations within the Pal\,5 stream, which had already been noticed by \citet{Odenkirchen03}. Here we are going to make use of such local density enhancements along the stream by exploiting the fact that these regions are the most likely places to find Pal\,5-like stars, and hence we will require our models to have high densities in these regions.\\\\  
To enhance a certain scale of density fluctuations, which are of the order of the stream width ($\approx0.2$\,deg, \citealt{Carlberg12b}), while  at the same time removing foreground/background variations and large-scale variations of the tail density itself, we adopted the \textit{Difference-of-Gaussians} process. This process is a simplification of the ``Laplacian of the Gaussian'' technique employed in computer vision for blob detection (e.g.~\citealt{Lindeberg98}). First, we masked out the central part of the cluster before convolving the map with the Gaussians. Then, we created a ``small-feature map'' by convolving the pixel values of the matched-filter, $\Sigma$, with a normal function, $\mathcal{N}$, of a certain width, $\sigma_1$, where $\sigma_1= 0.115$\,deg in our case (lower panel of Fig.~\ref{SDSS}). From this map we subtracted a ``background map'' smoothed with a larger kernel size, $\sigma_2 \approx 8\,\sigma_1$ (upper panel of Fig.~\ref{SDSS}). To compute the significances for each pixel in this ``residual map'', we followed \citet{Koposov08} and defined the significance, $S$, for each pixel to be 
\begin{equation}\label{eq:significance}
S = \sqrt{4\pi}\sigma_1\frac{\Sigma * \left(\mathcal{N}\left(\sigma_1\right)-\mathcal{N}\left(\sigma_2\right)\right)}{ 
\sqrt{\Sigma * \mathcal{N}\left(\sigma_2\right)}}.
\end{equation}
Fig.~\ref{significances} shows a histogram of significances, $S$, of the pixels of our residual map. As \citet{Koposov08} point out, the Poisson distribution of sources should yield a normal distribution for $S$ with a variance of 1. We consider pixels with values of $S$ larger than 3 as statistically significant (red part of the histogram in Fig.~\ref{significances}).\\\\
On this flat, background subtracted map, we searched for extended overdensities with \textsc{SExtractor} \citep{Bertin96}. The code was configured to search for groups of at least 10 connected pixels with $S>3$. To compute the FWHM and barycenters, the analysis threshold was lowered to $S>2$. Note that we masked the regions within $\approx1.5$\,deg of the borders of the SDSS footprint in order to avoid edge effects that arise from the \textit{Difference-of-Gaussians} process. Finally, we excluded overdensities found near the south-east edge of the footprint due to its proximity to the Galactic bulge (shaded region in Fig.~\ref{od_real}). This edge lies at $\approx 42\deg$ from the Galactic centre, which seems close enough to introduce significant noise in the matched filter.\\\\
The resulting 24 overdensities (19 in the trailing tail and 5 in the leading tail; shown in Fig.~\ref{od_real} and separated by a horizontal line) with a significance above $3\,\sigma$ are listed in Tab.~\ref{tab:overdensities}, sorted from large to small values of l\,$\cos\left(\mbox{b}\right)$. We also give their significances and FWHM as determined by \textsc{Sextractor}.\\\\ 
In our \textit{streakline} modeling, we will use the locations and widths of the detected blobs as the positions and positional uncertainties of the tidal tail centerline, or in other words as regions with highest local probability of finding Pal\,5 stars. We will use the overdensities to calculate a likelihood of each model (see Sec.~\ref{ssec:comparison}). In addition to these high-probability regions we will compare our \textit{streakline} models to the observed positions and radial velocities of stars in the tidal tails of Pal\,5.

\subsubsection{Radial velocities}

\begin{table}
 \centering
 \caption{Radial velocities of probable stream stars}
 \begin{tabular}{cccc}
 l\,$\cos(\mbox{b})$ [deg] & b [deg] & $V_R$ [km\,s$^{-1}$]& $\Delta V_R$ [km\,s$^{-1}$]\\
 \hline
 6.19 & 44.44 & -57.08 & 0.29 \\
 4.90 & 44.79 & -54.60 & 0.36  \\
 4.70 & 44.89 & -66.93 & 0.44  \\
 3.13 & 45.43 & -53.78 & 0.38  \\
 3.11 & 45.34 &-56.79  &0.32  \\
 3.05 & 45.25 &-51.83  &0.23  \\
 2.44 & 45.52 & -55.52 & 0.25 \\
 2.34 & 45.58 &-54.73  & 0.37 \\
\hline
 0.47 & 45.77 &-58.28  & 0.48 \\
-0.17 & 45.68 & -57.91 & 0.31 \\
-0.19 & 45.75 & -58.33 & 0.46 \\
-0.47 & 45.78 &-59.83  & 1.16 \\
-0.48 & 45.81  & -59.25 & 0.45 \\
-0.92 & 45.92 & -47.00 &  0.27 \\
-1.22 & 45.82 & -60.96 & 0.24 \\
-2.22 & 45.98 & -60.83 & 0.34 \\
-2.27 & 45.87 & -65.02 & 0.90 \\
 \end{tabular}
 \label{tab:radialvelocities}
\end{table}

Kinematic information on stars in a tidal stream can help significantly to constrain model parameters (see e.g.~\citealt{Pearson14}). The proper motion of Pal\,5 has to be assumed as being largely unknown, since three different photographic-plate measurements all yielded different results, which all disagreed within their stated uncertainties (\citealt{Schweitzer93, Scholz98}, and see discussion in \citealt{Odenkirchen01}). However, there is accurate information on its radial velocity and even on the radial velocities of some stars within its tidal tails. \\\\ 
\citet{Odenkirchen02} measured radial velocities of 17 giants within the tidal radius of Pal\,5 using VLT/UVES. From their analysis of these confirmed cluster members, they determined the heliocentric radial velocity of Pal\,5 with a very high precision to be $-58.7\pm0.2$\,km\,s$^{-1}$. This is much more precise than, e.g., the velocity uncertainty of the Solar motion with respect to the Galactic center. For this reason, we will neglect this uncertainty of 0.2\,km\,s$^{-1}$ on Pal\,5's radial velocity in the subsequent analysis.\\\\
In a later study, \citet{Odenkirchen09} presented VLT/UVES radial velocities of further 17 giants with locations well outside Pal\,5 but, in projection, close to the dense parts of its tidal tails. The locations and velocities of these stars are listed with their individual uncertainties in Tab.~\ref{tab:radialvelocities}. A horizontal line indicates, which stars lie in the trailing part of the tail (above), and which are in the leading (below).
Similar to the overdensities, we will compare our models to these 17 positions and velocities in order to compute a model's likelihood (see Sec.~\ref{ssec:comparison}). Note though that the locations of the overdensities represent the summarized positions of several stream stars, whereas the radial velocities represent individual stars in phase space that may not be representative of large fractions of stream stars.\\\\
Recently, \citet{Kuzma14} published 39 additional radial velocity measurements of red giants along the Pal\,5 stream. The radial velocity gradient along the Pal\,5 stream of their larger sample of $1.0\pm0.1$\,km\,s$^{-1}$\,deg$^{-1}$ matches the one determined by \citet{Odenkirchen09} of $1.0\pm0.4$\,km\,s$^{-1}$\,deg$^{-1}$. Such additional data will be valuable for future modeling of Pal\,5.

\subsection{Pal\,5-analog simulation data}\label{sssec:pal5analog}

In order to test our methodology of overdensity extraction and \textit{streakline} modeling, we ran an $N$-body model of a Pal\,5-like cluster. The parameters for this simulation were taken from the results of our first, preliminary modeling of the observations, and hence this analog $N$-body model of Pal\,5 is sufficiently close to reality (i.e.~our final fitting results) to allow such a comparison. For the integration of the model, we used the freely available, and GPU-enabled $N$-body code \textsc{Nbody6} \citep{Aarseth03, Nitadori12}, which we slightly modified such that it could integrate the cluster evolution in the same galactic potential as we used for our \textit{streakline} modeling of the Pal\,5 observations (see Sec.~\ref{ssec:potential} for a detailed description of the galactic potential).\\\\
In all our modeling, we kept the disk and bulge potential fixed as described in Sec.~\ref{ssec:potential}, since we mainly focussed on the potential of the dark halo. We assumed an NFW potential for the halo \citep{Navarro97}, which we allowed to be flattened along the z axis, i.e.~perpendicular to the galactic disk. For the analog model we chose a scale mass of $M_{Halo} = 1.475\times10^{12}\msun$, and a scale radius of $r_{Halo} = 36.5$\,kpc. The flattening was chosen to be $q_z  = 0.94$. The scale mass and scale radius correspond to a virial mass of $M_{200} = 1.60\times10^{12}\msun$, a virial radius of $r_{200} = 217$\,kpc, and a concentration of $c = 5.95$.\\\\
We set up the initial configuration for the star cluster model using the publicly available code \textsc{McLuster}\footnote{\url{https://github.com/ahwkuepper/mcluster.git}} \citep{Kupper11, mcluster}. For simplicity, we arbitrarily chose the initial cluster to be a Plummer sphere of 50\,000 single stars of 0.44$\msun$ each, with a half-mass radius of 11\,pc. This simple model is sophisticated enough to reproduce the shape of the tidal tails. We followed the evolution of the cluster for 6\,Gyr, during which it lost 10\,000$\msun$ of stars into the tidal tails, resulting in a final (i.e.~present-day) cluster mass of $\approx12\,000\msun$. Hence, the mass loss rate of this model is about 1.67$\msun$Myr$^{-1}$. The orbit of the cluster was chosen to be similar to our best-fit results from the observational study (cf.~Tab.~\ref{tab:orbits}), and yields an apocentric radius of 19.0\,kpc, a pericentric radius of 7.4\,kpc, and hence an orbital eccentricity of 0.44.\\\\
We then took a snapshot of the resulting distribution of stars from the location of the Sun (see Sec.~\ref{ssec:sun} for details) and inserted this stellar distribution into the SDSS source catalog with an offset of 4\,deg from the original Pal\,5. Each star particle from the simulation was randomly assigned a magnitude and color according to a stellar mass drawn from a present-day mass function \citep{Kroupa01}. These were generated such that the simulation particles roughly match Pal\,5's stellar population by using a PARSEC isochrone \citep{Bressan12} with an age of 11.5\,Gyr \citep{Dotter11} and a metallicity of [Fe/H] $= -1.3$ \citep{Smith02}.\\\\
It turned out that, when we used the full range of stellar masses down to $0.08\msun$, the number of stars in the tails of our Pal\,5-analog simulation bright enough to be within the SDSS magnitude limits is lower than the number of observed stars in the tails of the real Pal\,5. Thus, to get a similar number of stars in the simulated tidal tails as in the original Pal\,5 stream, we truncated the present-day mass function at the low-mass end of $0.5\msun$, so that enough stars fell into the SDSS magnitude limits. This may be interpreted as a first indication that the real Pal\,5 is losing stars at a significantly higher rate than our analog model, but could also originate from a different degree of orbital compression of the two streams. That is, if the real Pal\,5 is closer to apogalacticon or on a more eccentric orbit, the stream may be more strongly compressed and hence the surface density of stars in the tails may be higher.\\\\
On this new data set we applied the same methodology as described above, and extracted in total 23 new overdense regions for the analog stream. However, when we inserted the stream into the SDSS catalog, we had to shift it by 4\,deg so that it did not overlap with the real Pal\,5. Due to the angular shift, our analog stream was more affected by the Galactic bulge and ran earlier off the SDSS footprint. Hence, the analog model did not produce statistically significant overdensities at the far end of the trailing tail and is therefore 4 deg shorter in one direction. The leading tail, on the other hand, extends about 2 more degrees than the observed Pal\,5. Thus, the two streams, observed and simulated, have a similar length and a comparable number of overdense regions.\\\\
We also generated a set of mock radial velocities from the simulation. From the set of $N$-body particles we randomly picked 17 particles with similar locations on the sky as the 17 giants \citet{Odenkirchen09} observed. As uncertainty for each radial velocity we chose a conservative value of 0.5\,km\,s$^{-1}$ (cf.~Tab.~\ref{tab:radialvelocities}). These are the ingredients that went into our \textit{streakline} modeling. In the following we are going to describe the details of this modeling process.

\subsection{Streakline modeling}\label{ssec:streaklines}

Ideally, we would like to create a comprehensive set of direct $N$-body models of Pal\,5 and compare them to the available set of observations. However, even for such a sparse and low-mass globular cluster like Pal\,5, a full $N$-body computation of the cluster and its forming tidal tails, covering several Gyr of cluster evolution, takes a few days on a modern GPU workstation with the high-end code \textsc{Nbody6}. A parameter-space study involving many hundred thousand realizations of Pal\,5 in different Galactic potentials and on different orbits, like we envisage here, is out of question with such an approach. Therefore, we will use \textit{streakline} models to generate realistic globular cluster stream realizations within a matter of seconds, where we follow the \texttt{FAST-FORWARD} modeling approach developed in \citet{Bonaca14}. The concept of this method is described in the following.

\subsubsection{Escape of stars from a cluster}\label{sssec:escape}

\citet{Kupper12} showed that the formation of tidal tails of Pal\,5-like clusters on typical orbits within a galaxy can be very well understood and approximated by restricted three-body simulations in which massless test particles are integrated together with a massive cluster particle within a background galaxy potential. \citet{Kupper12} demonstrated that cluster stars preferentially escape through the Lagrange points of the cluster-galaxy system ($\vec{R_L}$), which can be calculated to lie at a distance of
\begin{equation}\label{eq:tidalradius}
r_{L} = \left( \frac{G\,M_c}{\Omega_c^2-\partial^2\Phi/\partial R_c^2}\right)^{1/3}
\end{equation}
from the cluster center along the line connecting the galactic center and the cluster center \citep{King62}. In this equation, $G$ is the gravitational constant, $M_c$ is the cluster mass, $\Omega_c$ is the instantaneous angular velocity of the cluster on its orbit within the galaxy, and $\partial^2\Phi/\partial R_c^2$ is the second derivative of the galactic potential, $\Phi$, with respect to the galactocentric radius of the cluster, $R_c$. Moreover, \citet{Kupper12} showed that for low-mass satellites as Pal\,5 the velocities of escaping stars, $\vec{V}$, are preferentially such that they match the angular velocity of the cluster center with respect to the galactic center. That is, we can generate the initial conditions for the test particles ($\vec{R}$, $\vec{V}$) by using
\begin{eqnarray}
	R^i &=& \frac{R^i_c}{R_c} \times \left( R_c\mp r_L\right)\mp\delta r^i,\\
       V^i &=& \frac{V^i_c}{V_c} \times \left(V_c\mp \Omega_c\,r_L\right)\mp \delta v^i,
\end{eqnarray}
with $i$ denoting the x, y, and z components, and the upper sign being for the leading tail while the lower one is for the trailing tail. The two random offsets, $\vec{\delta r}$ and $\vec{\delta v}$, can be added to the initial conditions of the test particles to emulate the scatter around these idealized escape conditions as seen in full $N$-body simulations \citep{Kupper12, Lane12, Bonaca14, Amorisco14, Gibbons14, Fardal14}.\\\\
\citet{Bonaca14} demonstrated that with a realistic choice for the scatter in the escape conditions it is possible to recover the underlying model parameters of direct $N$-body simulations of a dissolving globular cluster in an analytic galaxy potential. Here we chose a slightly different approach from the \texttt{FAST-FORWARD} modeling of \citet{Bonaca14} by setting these random components to be zero in order to create the kinematically coldest stream possible. We do this since we are interested in reproducing the epicyclic substructure in the stream with as few test particles as possible (to save computational time), and this substructure will be preferentially produced by the coldest escapers \citep{Kupper12}. Moreover, \citet{Kupper08b} and \citet{Just09} showed that most escapers will evaporate from a star cluster with just about the right amount of energy to escape. This is the family of escapers that will produce the substructure in the stream, whereas higher-temperature escapers or ejecters (e.g.~from three-body interactions) will basically only introduce noise in the substructure pattern or not orbit within the tails at all. In Sec.~\ref{sssec:pal5_sim} we will demonstrate that this approach recovers the model parameters of direct $N$-body simulations of a dissolving globular cluster on a Pal\,5-like orbit perfectly well.

\subsubsection{Recapture of stars and the edge radius}\label{sssec:edgeradius}

This picture of the escape process, which we use to generate our \textit{streakline} models, is somewhat different from the notion introduced by \citet{King62}, who assumed that cluster stars are stripped at perigalacticon such that the limiting radii of the globular clusters we observe today correspond to their tidal radii (Eq.~\ref{eq:tidalradius}) at the perigalactica of their orbits. However, it was shown with $N$-body simulations in \citet{Kupper10a} and \citet{Webb13} that clusters adjust to the mean tidal field of an orbit rather than to the perigalactic tidal field. \citet{Kupper12} demonstrated that the tidal radius according to King's formula for a cluster on an eccentric orbit gets so small at perigalacticon that stars cannot escape from there\footnote{Unless they have high excess energies, e.g.~from three-body encounters or very strong tidal shocks.}. Instead, they will get recaptured by the star cluster while it is moving into apogalacticon, as the tidal radius quickly grows to its maximum value. This demonstrates the importance of the cluster's gravitational acceleration on the trajectories of escaping stars. In fact, \citet{Kupper12} showed that, even for clusters on circular orbits, the trajectories of tail stars are significantly affected by the cluster mass. This gets more important for more eccentric orbits, since tail stars will have periodic strong encounters with the cluster in apocenter. For very eccentric orbits, this effect even influences the shape of the tidal tails at large distances from the cluster, as the compression of the tidal tails in apocenter is so significant that distant debris is pushed back into the cluster vicinity and remains close to the cluster for a significant fraction of the orbital period.\\\\
\citet{Kupper12} suggest to introduce a lower limit on the tidal radius from Eq.~\ref{eq:tidalradius} in order to prevent recapture of test particles. In our investigation we chose this \textit{edge radius} to be at least equal to Pal\,5's half-light radius of 20\,pc \citep{Harris96}, and iteratively increase it for a specific test particle if it gets recaptured by the cluster. The cluster itself is represented by a softened point mass, $M_c$, with a softening length of the order of the cluster's half-light radius of 20 pc. The influence of this latter number on the actual shape of the tails is negligible, which is why we fixed it to some reasonable value instead of making it a free parameter in the modeling process. For more massive and much more extended systems like the Sagittarius dwarf galaxy, the progenitor should be modeled with more care as the extent of the system more strongly influences the shape of the tidal tails \citep{Gibbons14}.

\subsubsection{Test-particle integration}

The \textit{streakline} models are generated in an analytic background galaxy potential, $\Phi$, which will be specified in the next Section. Given the more or less well constrained 6D phase-space coordinates of Pal\,5 at the present day, we integrate the cluster particle from this initial position backwards in time for a specified integration time, $t_{int}$, using a 4th-order Runge-Kutta integrator. From this point in the past ($-t_{int}$) we integrate back to the future until we reach the present day, while releasing test particles at a fixed time interval, $dt_{release}$. Once a test particle is released from the cluster particle at a given point in time in the past, the test particle is integrated together with the cluster particle within the galaxy potential until the present day. This is repeated $K = t_{int}/dt_{release}$ times until the cluster particle has reached its present-day position. The test particles do not interact with each other, as the inter-particle distance within the stream is too large to cause relevant interactions. Thus, each \textit{streakline} model is a combination of $K$ restricted three-body integrations, where the longest integrations are over a time interval $t_{int}$ and the shortest ones are only of length $dt_{release}$. We chose a interval length of $dt_{release} = 10$\,Myr, which we found to produce well-resolved stream models with acceptable computational effort. Pal\,5 loses stars at a much higher rate, though (probably of the order of 20 stars per Myr, when we use our fitting result of $7.9\msun$\,Gyr$^{-1}$ and assume a mean stellar mass of $0.4\msun$). Our test particles are therefore only tracers of the stream, not exact representations of the actual stellar densities.\\\\
Note that this approach produces streaklines with a constant release rate of test particles. Tidal shocks, e.g.~when the cluster crosses the galactic disk, do not increase the release rate temporarily. This appears to be justified as no obvious density variations resulting from disk shocks can be found in $N$-body simulations of star clusters on Pal\,5-like orbits \citep{Dehnen04, Kupper10b}. For real clusters and full $N$-body simulations of star clusters, the eccentricity of an orbit \textit{does} change the mass-loss rate, though, as higher eccentricities will make the cluster plunge deeper into the central parts of the galactic potential \citep{Baumgardt03}. However, this increased mass loss appears to be spread out over the entire orbit of the cluster \citep{Kupper10b}. This delayed escape of stars that got unbound during a tidal shock is due to the relatively long timescale of escape for just slightly unbound stars \citep{Fukushige00, Zotos15}, as well as due to the recapture of many stars that got temporarily unbound during a tidal shock as mentioned above (Sec.~\ref{sssec:edgeradius}). Of course, stars that gained a lot of energy during a tidal shock will quickly escape from the cluster. This can be seen for example in simulations of dwarf galaxies falling radially into the potential of a host galaxy. In such extreme cases the tidal shocks will produce substructure in the satellite's stream \citep{Penarrubia09}. However, such stars with high excess energies will not contribute to the coldest part of the stream that we are interested in. Our \textit{streakline} method therefore produces cold tidal streams, which trace the density variations along the streams due to orbital compression and stretching of the stream in apogalacticon and perigalacticon, respectively, as well as due to epicyclic motion. But the density of test particles does not directly correspond to the density of stars in the stream, as the fraction and distribution of kinematically warmer stream particles is not well known. The test particles are therefore just tracers of the locations where the coldest stream stars pile up.\\\\
Similar approaches of test-particle integrations for tidal streams have been used in previous studies \citep{Odenkirchen03, Varghese11, Gibbons14}, but in our case the method is tailored to globular clusters and calibrated with $N$-body simulations, which makes the resulting stream models more realistic and -- as we will show -- estimates on, e.g., parameters of the Galactic potential more reliable.

\subsection{Galactic potential}\label{ssec:potential}

The gravitational potential of the Milky Way (MW) is one of the main quantities of interest here. The choice of representation or, in most cases, analytical parametrization of the potential is not straightforward, and ideally it should be kept as flexible as possible. Great flexibility, however, implies many degrees of freedom. Given the limited amount of data we currently have for Pal\,5, we want to keep the free parameters at a minimum but still be flexible enough to accurately describe the MW potential and get a meaningful comparison to other studies.\\\\
In past studies, the Galactic potential has often been parametrized as a three-component model consisting of a bulge, a disk and a halo (e.g.~\citealt{Johnston99, Helmi04a, Johnston05, Fellhauer07, Koposov10, Law10}). We will adopt a similar parametrization and focus on the halo component, as this is the least constrained, but also the most massive component of the Milky Way. The other two components, bulge and disk, are kept fixed and are modeled as follows. For the bulge potential we use a Hernquist spheroid \citep{Hernquist90},
\begin{equation}
\Phi_{Bulge} (r) = -\frac{GM_{Bulge}}{r+a},
\end{equation}
where $G$ is the gravitational constant, $r = \sqrt{x^2+y^2+z^2}$ is the Galactocentric radius, $M_{Bulge} = 3.4 \times 10^{10} \msun$, and  $a$ = 0.7\,kpc. As disk component we use an axis-symmetric \citet{Miyamoto75} potential,
\begin{equation}
\Phi_{Disk} (R,z) = - \frac{GM_{Disk}}{\sqrt{R^2+\left(b + \sqrt{z^2+c^2}\right)^2}}
\end{equation}
with $R = \sqrt{x^2+y^2}$, $M_{Disk} = 10^{11} \msun$, $b$ = 6.5\,kpc, and $c$ = 0.26\,kpc. These parametrizations and also these parameter values for the two stellar components are not the most sophisticated or up-to-date choices (see e.g.~\citealt{Dehnen98, Binney08, Bovy13, Irrgang13, Licquia14}), but we chose them here for comparability to past studies. In follow-up studies with more observational data for Pal\,5, and when other constraints on the Galactic potential are taken into account, the baryonic components should be included in the modeling process. However, we argue that the differences between the various parametrizations of these components at the average Galactocentric distance of Pal\,5 are not very important, as the halo is the dominant mass component at these radii. Moreover, \citet{Vesperini97} showed that the Galactic disk has very little effect on the evolution of globular clusters at Galactocentric radii beyond about 8 kpc, which is where Pal\,5 crosses the disk most of the time.\\\\
We will use these analytically simple and computationally inexpensive prescriptions for the bulge and disk, and focus on varying the mass, shape and extent of the dark halo. As parametrization for the halo we chose the NFW density profile \citep{Navarro97}, as it has been shown to be a very good representation for dark matter halos in $\Lambda$CDM simulations of structure formation (see e.g.~\citealt{Bonaca14}). The gravitational potential resulting from this density profile has the form
\begin{equation}\label{eq:halo}
\Phi_{Halo} (r) = -\frac{G\,M_{Halo}}{r} \ln\left(1+\frac{r}{r_{Halo}}\right),  
\end{equation}
where $M_{Halo}$ and $r_{Halo}$ are a scale mass and a scale radius, respectively, which are left as free parameters in our modeling. The scale mass and scale radius can be easily related to the more frequently used virial radius, $r_{200}$, the virial mass, $M_{200}$, and the concentration, $c$. $r_{200}$ is the radius of a sphere with a mean interior density of 200 times the critical density, $\rho_{crit}$, which has a value of $\rho_{crit} = 3\,H^2/(8\pi\,G) = 1.26\times 10^{-7}\msun$\,pc$^{-3}$ when using a Hubble constant of $H = 67.3$\,km\,s$^{-1}$Mpc$^{-1}$ \citep{Planck13}. The virial mass, that is the mass enclosed in a spherical volume of radius $r_{200}$, is related to the scale mass via
\begin{equation}
\frac{M_{200}}{M_{Halo}} = \left(\frac{r_{Halo}}{r_{Halo} + r_{200}} + \ln\left(\frac{r_{Halo} + r_{200}}{r_{Halo}}\right) - 1.0\right),
\end{equation}
and the concentration of the NFW halo is usually defined as $c = r_{200}/r_{Halo}$.\\\\
For computational convenience we will use Eq.~\ref{eq:halo} for our modeling, but report our results also in terms of $r_{200}$, $M_{200}$ and $c$ to allow comparability to other studies. However, due to the flattening along the z-axis, and due to the vast extrapolation from Pal\,5's apogalactic radius of about 19\,kpc out to a virial radius of about 200-300\,kpc, which solely relies on the assumption of an NFW profile for the halo, comparisons between these numbers from different studies have to be taken with care.\\\\
This is especially important since we also allow the potential to be flattened along the z-axis perpendicular to the Galactic disk by introducing a scale parameter $q_z$. That is, in the above equation for the halo potential we replace $r$ by $\sqrt{R^2+z^2/q_z^2}$. Such a potential flattening makes the definition of an enclosed mass ill-defined. A more reasonable quantity, which our modeling may provide, is the acceleration Pal\,5 experiences at its present-day position high above the Galactic disk, $\vec{a}_{Pal\,5}$. This acceleration can be converted into a local ``circular velocity'' 
\begin{equation}\label{eq:vcpal5}
V_C \left(r_{Pal\,5}\right) = \sqrt{\left|a_{Pal\,5}\right|\,r_{Pal\,5}},
\end{equation}
and as such allows a convenient comparison to the gravitational field within the Galactic disk, for which accurate rotation curves have been measured (e.g.~\citealt{Sofue13}). We will therefore calculate these values from our modeling results, and also compute the circular velocity at the Solar Circle to compare the performance of our modeling with results from data sets independent of Pal\,5.

\subsection{Solar parameters}\label{ssec:sun}

One key ingredient for the comparison of model points from numerical simulations to data points from observations is the location and motion of the Sun within the Galaxy. We chose our right-handed Galactocentric coordinate system such that the Sun lies in the xy-plane on the x-axis with the x-axis pointing to the Galactic center, and such that its tangential motion is along the y-axis in the direction of positive y.\\\\
Observations of the, so-called, S-stars around the Galactic center suggest that the distance of the Sun from the Galactic center is $R_{\odot} = 8.33\pm0.35$\,kpc \citep{Gillessen09}, which puts it in our coordinate system to $\vec{R}_\odot = (-8.33, 0, 0)$\,kpc. Using parallaxes and proper motions of over 100 masers in the Milky Way, \citet{Reid14} found a similar but somewhat more precise value of $8.34\pm0.16$\,kpc. With this data the authors were also able to constrain the tangential motion of the Sun within the Galaxy. This motion is usually split into the circular velocity of the Galaxy at the Solar radius, $V_C$, and the Solar reflex motion ($U_\odot, V_\odot, W_\odot$), that is, the motion of the Sun with respect to an object on a closed circular orbit at $R_\odot$. \citet{Reid14} find the total tangential velocity of the Sun in y-direction to be $V_{tan} = V_{C} + V_\odot = 255.2\pm5.1$\,km\,s$^{-1}$.\\\\
\citet{Schonrich12} used Hipparcos data to derive the Sun's reflex motion to be
\begin{equation}\label{eq:solarmotion}
\vec{V}_{\odot} = \left(11.1\pm0.7, 12.2\pm0.5, 7.3\pm0.4\right)\,\mbox{km\,s}^{-1},
\end{equation}
which gives us a value of 243\,km\,s$^{-1}$ for the circular velocity at the Solar radius. A similar result of $V_{C} = 239 \pm 5$\,km\,s$^{-1}$ was found independently by \citet{McMillan11} using a wider range of observational data.\\\\
However, from APOGEE data, \citet{Bovy12} determined a slightly different set of Solar parameters, putting the Sun at a distance of $8.1^{+1.2}_{-0.1}$\,kpc from the Galactic Center with a total tangential velocity of $V_{tan} = 242^{+10}_{-3}$\,km\,s$^{-1}$, only marginally in agreement with the \citet{Reid14} values. Moreover, while their $U_\odot$ and $W_\odot$ components are in good agreement with the \citet{Schonrich12} values, they determined the circular velocity at the Solar Circle to be as low as $218\pm6$\,km\,s$^{-1}$.\\\\
Due to this existing discrepancy, we will not use literature constraints on the circular velocity at the location of the Sun within the Milky Way nor on its tangential velocity. For the coordinate transformations between observer coordinates and the Cartesian coordinates of the models we will adopt the \citet{Schonrich12} values for $U_\odot$ and $W_\odot$, but we will keep $V_{tan}$, as a free parameter. Moreover, we will leave the distance of the Sun from the Galactic Center as a free parameter.

\subsection{Comparison of models to observations}\label{ssec:comparison}
The most important and also most sensitive part for drawing conclusions from the modeling of tidal streams is the comparison of the models to the observational data. In the following we will use the formalism outlined in \citet{Bonaca14}, which is based on the framework developed in \citet{Hogg10} and \citet{Hogg12}.\\\\
\citet{Bonaca14} defined a likelihood, $P$, for a given set of model parameters, $\theta$, and priors, $I$, to compare stream particles from the \textit{Cauda} simulation (a re-simulation of the \textit{Via Lactea II} simulation by \citealt{Diemand08} including globular cluster streams) to particles from \textit{streakline} models. Similar to this work, we will refer to our \textit{streakline} particles as model points, $x_k$, while our data points will be the Pal\,5 observations or the Pal\,5-analog observations, respectively, denoted as $X_n$.\\\\
Assume we have $K$ model points and $N$ data points. We can write the probability, $p$, of a model point, $x_k$, generating a data point, $X_n$, as
\begin{equation}
p\left(X_n | x_k, \theta, I\right) = \mathcal{N}\left( X_n | x_k, \sigma_n^2 +\Sigma^2_k\right),
\end{equation}
where $\theta$ is the set of model parameters that generated the model point, and $I$ is any prior knowledge we have on the parameters. The model parameters $\theta$ and priors $I$ will be described in detail in Sec.~\ref{ssec:MCMC}. We assess the probability of each model point to generate a data point by either using a 2-dimensional or 3-dimensional Gaussian distribution, $\mathcal{N}\left(X | m, V\right)$, for $X$ with mean, $m$, and variance tensor, $V$, for the overdensities or for the radial velocity stars, respectively. That is, for the overdensities, the two dimensions are Galactic longitude and Galactic latitude, whereas for the radial velocity stars we add the velocity as a third dimension.\\\\
The variance tensor $V$ consists of the observational uncertainty tensor, $\sigma_n^2$, and a smoothing tensor, $\Sigma_k^2$. For the overdensities we will use their FWHM as positional uncertainty, whereas for the radial velocity stars we use a positional uncertainty of 0.115\,deg, which accounts for the finite width of the stream. Moreover, we use their individual uncertainties in velocity from the observations as given in \citet{Odenkirchen09} and listed in Tab.~\ref{tab:radialvelocities}. The tensor $\Sigma_k^2$ allows us to smooth the discrete distribution of points such that their Gaussians overlap in phase space. This makes the generative model of the stream continuous in phase space, i.e.~it turns the discrete streakline particles into a probability density function in phase space with finite support everywhere. The smoothing length in each dimension is kept as a free hyper-parameter in $\theta$, over which we will marginalize in the end before we report the results for the other parameters.\\\\
For each data point, $X_n$, we can determine the likelihood of being produced by the whole \textit{streakline} model by marginalizing over the $K$ model points
\begin{equation}
p\left(X_n | \theta, I\right) = \sum_{k=1}^K P_k\,p\left(X_n | k, \theta, I\right),
\end{equation}
with the normalization factor being $P_k = 1/K$. The number of model points $K$ is in principle an arbitrary choice, and hence the likelihood should not depend on this number. A model with higher number of particles will sample the density distribution of the tidal tails better and will reduce noise. However, producing the model points is the computationally expensive part of the whole modeling process, so it should be kept at a reasonable number. \citet{Bonaca14} find that for a computationally efficient scan of the multi-dimensional parameter space of the \textit{streakline} models, we should use $K$ being a few times larger than $N$. While \citet{Bonaca14} use a fixed factor of 5 difference, we will use a variable number of model points and different setups for the number of data points. However, in any case we will keep $K/N>3$, and in most cases it is between 5 and 20. This variability in $K$ comes from our choice of keeping the interval, at which model points are created, fixed to $dt_{release} = 10$\,Myr, while leaving the integration time, $t_{int}$, as a free hyper-parameter (see next Section).\\\\
To compute the combined likelihood of the set of model points creating the set of data points, we assume that all data points are independent of each other, so that we can multiply their individual likelihoods to yield 
\begin{equation}\label{eq:likelihood}
P \left( \{X_n\} | \theta, I\right) = \prod^N_{n=1}p\left(X_n | \theta, I\right).
\end{equation}
This likelihood will be used as input for the MCMC algorithm described in the next Section. Similar applications of this formalism can be found in \citet{Bonaca14} and \citet{Price14}.

\subsection{Markov chain Monte Carlo (MCMC) machinery}\label{ssec:MCMC}

\begin{table}
\label{tab:parameters}
\centering
\caption{Model parameters and range of flat priors}
\begin{tabular}{l|cl}
Parameter & Prior range & Description\\\hline
$M_{Halo}$ & $[10^9;10^{13}]\msun$ &Halo scale mass  \\
$r_{Halo}$ & $[0.1;100]$\,kpc & Halo scale radius\\
$q_{z}$ & [0.2;1.8] & Halo flattening  \\\hline
$M_{Pal\,5}$ &  $[0; 60000]\msun$ & Cluster present-day mass\\
$dM/dt$ & $[0;100]\msun$\,yr$^{-1}$  &  Average cluster mass-loss rate   \\
$d_{Pal\,5}$ & $[20;27]$\,kpc & Cluster distance from Sun \\
$\mu_{\alpha}\cos(\delta)$ &  $[-3;-1]$\,mas\,yr$^{-1}$ & Proper motion in RA\\
$\mu_{\delta}$ & $[-3;-1]$\,mas\,yr$^{-1}$  & Proper motion in Dec\\\hline
$V_{tan}$ & $[212;292]$\,km\,s$^{-1}$ & Solar transverse velocity\\
$R_{\odot}$  & $[7.5;9.0]$\,kpc & Solar Galactocentric distance\\
\end{tabular}
\end{table}

\begin{table}
\label{tab:coefficients}
\centering
\caption{Polynomial coefficients for stream centerline}
\begin{tabular}{l|cc}
Coefficient & Simulation & Observations \\
\hline
$a_0$ & $45.9373$ & $45.9816$ \\
$a_1$ & $-6.13919\times 10^{-2}$ & $-2.50988\times 10^{-2}$ \\
$a_2$ & $-1.3957\times 10^{-2}$ & $-2.44554\times 10^{-2}$ \\
$a_3$ & $-1.96972\times 10^{-4}$ & $3.32027\times 10^{-4}$ \\
\end{tabular}
\end{table}

Using Eq.~\ref{eq:likelihood} we can assess the likelihood of a given \textit{streakline} model to produce the observational data. This likelihood we can feed into the Markov chain Monte Carlo code \texttt{emcee} \citep{Foreman13} to find the posterior probability density for the model parameters. \texttt{emcee} is an affine-invariant ensemble sampler, which is ideal for massively parallel runs on high-performance computing clusters using message-passing interfaces such as OpenMP and MPI (e.g., \citealt{Allison14}). Moreover, it can make use of parallel tempering, which allows us to scan the parameter space at different ``temperatures'' (e.g., \citealt{Varghese11}). That is, the chains will basically walk through parameter space with different stride lengths. Parallel tempering is useful for multi-modal likelihood surfaces, as it helps walkers to move out of local maxima, and hence enables a more efficient scan of the parameter space.\\\\
In total we have 10 free model parameters and 4 hyper-parameters. The model parameters can be grouped into three categories:
\begin{enumerate}
\item halo parameters (scale mass, $M_{Halo}$, scale radius, $r_{Halo}$, flattening, $q_z$), 
\item cluster parameters (present-day mass, $M_{Pal\,5}$, average mass-loss rate, $dM/dt$, distance, $d_{Pal\,5}$, 2 proper motion components, $\mu_{\alpha\cos(\delta)}$ and $\mu_\delta$),
\item Solar parameters (transverse velocity with respect to the Galactic Center, $V_{tan}$, and Galactocentric radius, $R_\odot$).
\end{enumerate}
We set up the model parameters in a very wide range around possibly allowed values, which we roughly constrain with values from the literature. We will use these same bounds for the initial values also as (flat) priors, that is, if a walker tries to ``walk'' outside this bound, the likelihood is set to $-\infty$. The parameters and the range of initial values/the priors are listed in Tab.~\ref{tab:parameters}.\\\\
Our four hyper-parameters are: the smoothing in position for the overdensities, the smoothing in position for the radial velocity stars, and the smoothing in velocity for the radial velocity stars. The fourth hyper-parameter is the integration time for the \textit{streakline} models. The latter has to be kept as a free nuisance parameter since we do not know a priori how old the part of the Pal\,5 stream is that we see within the SDSS footprint. It may well be that the stream extends much further in both leading and trailing direction, or that large parts of the tails have dispersed over time. Hence, we allow integration times between 2 and 10 Gyr, and smoothings between 0 and 100\,deg or km\,s$^{-1}$, respectively.\\\\  
We ran \texttt{emcee} on 128 cores of the Yeti cluster at Columbia University using a setup of two temperatures, with 512 walkers in each temperature. After a ``burn-in'' phase of 200 steps per walker, we checked that the chains had converged sufficiently and started the sampling phase. For the final posterior probability densities, which we are going to report in the following, we ran the chains long enough such that each walker had made between 500 and 1000 steps. In the following, we are going to report the results from the lowest temperature chains.\\\\
Since the overdensities and the radial velocity stars are different type of data, and since it is not straightforward how to combine them into one statistic, we ran 4 different analysis approaches on both the analog simulation and on the Pal\,5 observations. The 4 different configurations are the following:
\begin{enumerate}
\item \textit{Overdensities}: we assess the likelihood of the model producing the observational data by only using the locations of the overdensities as described in Sec.~\ref{ssec:overdensities} and listed in Tab.~\ref{tab:overdensities} for the observations. All overdensities are treated the same, and their FWHM are treated as the uncertainties of their barycenter positions.
\item \textit{Overdensities + radial velocities}: in addition to the locations of the overdensities, we add the locations of the radial velocity stars in galactic longitude, galactic latitude and radial velocity. 
\item \textit{Weighted overdensities}: using the FWHM as positional uncertainties of the overdensities is most probably a considerate overestimate. In an attempt to assign more weight to the more pronounced (and more extended) overdensities, we use their significances, $S$, as weight factors. The likelihood function (Eq.~\ref{eq:likelihood}) then reads
\begin{equation}
P \left( \{X_n\} | \theta, I\right) = \prod^N_{n=1}p\left(X_n | \theta, I\right)^S.
\end{equation}
We assign the radial velocity stars a significance of 1, whereas the overdensities have significances ranging from 3 to 10 (Tab.~\ref{tab:overdensities}). This weighting of the overdensities compared to the radial velocity stars is somewhat arbitrary, but as it turns out the actual choice does not have a significant effect on the results. The sum of all significances of the 24 overdensities from the Pal\,5 observations is 117 (136 for the 23 overdensities used for the analog simulation), which means that we effectively upweight our SDSS data by a factor of 5 compared to the radial velocity data. However, this up-weighting is not linearly distributed over the overdensities, but according to their observed significances. Thus, an overdensity that is detected, e.g., with a significance of 3 and a FWHM of 0.1\,deg counts as if we detected three Pal\,5 stars at the location of the overdensity's barycenter with each having an uncertainty of 0.1\,deg. 
\item \textit{Interpolated centerline}: as an alternative approach, we interpret the locations of the overdensities as the centerline (or centroid) of the stream, and interpolate points between these observed overdensities. Given Pal\,5's convenient shape in Galactic longitude, $l$, and latitude, $b$, we fit a third-order polynomial of form
\begin{equation}
b(l) = a_0 + a_1\,l + a_2\,l^2 + a_3\,l^3
\end{equation}
to the data points using a Marquardt-Levenberg algorithm. The resulting coefficients for the analog simulation and for the observations are listed in Tab.~\ref{tab:coefficients}. With this polynomial we generate 117 points for the observations and 136 points for the simulation along the whole extent of the stream. We chose these points to be equally spaced in Galactic longitude with a spacing width of 0.18\,deg (0.13\,deg). The positional uncertainties of these centerline points we set to be twice the spacing width, such that we have a continuous representation of the stream along this fitted centerline. We excluded the parts of the centerline between the innermost overdensities, i.e.~T1 and L1 (see Tab.~\ref{tab:overdensities}). In this way, we can use our Bayesian formalism of Sec.~\ref{ssec:comparison} and directly compare the results of this centerline fitting to the other approaches.
\end{enumerate}
Each of the four analysis approaches was run on both datasets, the analog simulation and the real observations of Pal\,5. As mentioned above, we used an MCMC algorithm to explore the model parameter space and determine the posterior probability densities of each parameter. These densities were determined from the steps of the walkers in the Markov chains. We used 512 walkers, and each of them made at least 500 steps, resulting in >256000 steps through parameter space. These steps provide us with individual histograms for each model parameter, showing us how often the chains have used a certain value for a given parameter. We treat as our best-fit parameter the median value of the posterior distribution for a parameter, and use the values containing 68\% of choices above and below the median as the upper and lower uncertainty range, respectively.

\section{Results}\label{sec:results}
In the next sections, we are going to compare the four different analysis approaches described above. First we will do this for the analog simulation, which will give us a measure of how well the methods recover the simulation parameters, i.e.~accuracy and precision, and then we will report the results for the Pal\,5 observations.

\subsection{Pal\,5-analog simulation}\label{ssec:resultsanalog}

\begin{table*}
\label{tab:accuracies}
\centering
\caption{Accuracy from fitting to the simulation}
\begin{tabular}{l|cccc}
 & Overdensities & Overdensities & Weighted & Interpolated\\
Parameter                  &                       &  + radial velocities &   overdensities  & centerline\\ \hline
$M_{Halo}$ & $1.19^{+0.52}_{-0.43}$  & $1.17^{+0.56}_{-0.46}$   & $1.02^{+0.31}_{-0.28}$  & $1.08^{+0.36}_{-0.27}$   \\
$r_{Halo}$ & $0.85^{+0.59}_{-0.37}$  & $1.16^{+0.35}_{-0.30}$   & $1.10^{+0.25}_{-0.20}$  & $1.17^{+0.23}_{-0.19}$   \\
$q_{z}$ & $1.05^{+0.25}_{-0.28}$  & $1.10^{+0.27}_{-0.28}$   & $1.11^{+0.16}_{-0.16}$  & $1.18^{+0.16}_{-0.17}$   \\\hline
$M_{Pal\,5}$ & $1.17^{+1.23}_{-0.83}$  & $1.15^{+0.83}_{-0.72}$   & $1.19^{+0.50}_{-0.57}$  & $0.41^{+0.50}_{-0.24}$   \\
$dM/dt$ & $2.32^{+3.53}_{-1.85}$  & $2.54^{+4.30}_{-2.19}$   & $1.03^{+4.75}_{-0.70}$  & $0.97^{+0.90}_{-0.57}$   \\
$d_{Pal\,5}$ & $1.00^{+0.05}_{-0.05}$  & $1.00^{+0.05}_{-0.05}$   & $1.00^{+0.04}_{-0.04}$  & $0.99^{+0.03}_{-0.04}$   \\
$\mu_{\alpha}\cos(\delta)$ & $1.03^{+0.12}_{-0.11}$  & $0.99^{+0.09}_{-0.07}$   & $1.00^{+0.07}_{-0.07}$  & $1.00^{+0.06}_{-0.05}$   \\
$\mu_{\delta}$ & $1.03^{+0.10}_{-0.11}$  & $0.98^{+0.09}_{-0.07}$   & $1.00^{+0.07}_{-0.07}$  & $1.01^{+0.06}_{-0.05}$   \\\hline
$V_{tan}$ & $1.00^{+0.10}_{-0.10}$  & $0.99^{+0.08}_{-0.08}$   & $1.03^{+0.07}_{-0.07}$  & $1.03^{+0.07}_{-0.07}$   \\
$R_{\odot}$  & $1.00^{+0.05}_{-0.05}$  & $0.99^{+0.04}_{-0.04}$   & $1.00^{+0.03}_{-0.03}$  & $1.03^{+0.03}_{-0.03}$   \\\hline
$V_{C}(R_\odot)$ & $1.08^{+0.24}_{-0.13}$  & $0.98^{+0.06}_{-0.04}$   & $0.98^{+0.04}_{-0.03}$  & $0.98^{+0.03}_{-0.02}$   \\
$a_{Pal\,5}$ & $1.30^{+0.84}_{-0.49}$  & $0.93^{+0.25}_{-0.17}$   & $0.90^{+0.18}_{-0.13}$  & $0.90^{+0.14}_{-0.11}$   \\
$M_{200}$ & $1.41^{+0.92}_{-0.70}$  & $1.05^{+0.51}_{-0.41}$   & $0.92^{+0.32}_{-0.26}$  & $0.95^{+0.30}_{-0.25}$   \\
$r_{200}$ & $1.12^{+0.20}_{-0.23}$  & $1.02^{+0.14}_{-0.16}$   & $0.97^{+0.10}_{-0.10}$  & $0.98^{+0.09}_{-0.09}$   \\
$c$ & $1.32^{+1.19}_{-0.64}$  & $0.88^{+0.30}_{-0.21}$   & $0.89^{+0.20}_{-0.18}$  & $0.84^{+0.16}_{-0.13}$   \\
\end{tabular}
\end{table*}

\begin{figure*}
\centering
\includegraphics[width=0.9\textwidth]{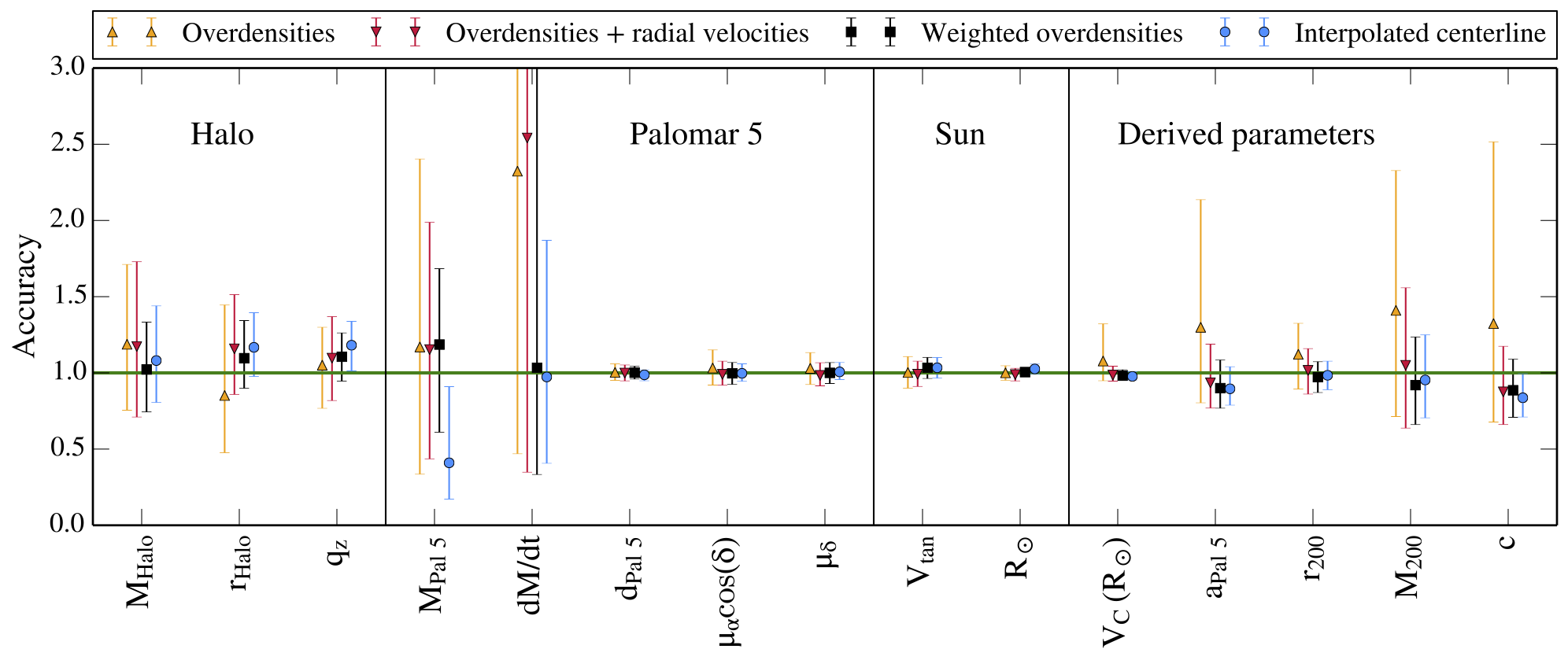}
\caption{Summary of the accuracies in reproducing parameters from the analog simulation for the four different modeling approaches: using only the overdensities, using the overdensities plus the radial velocities, using the weighted overdensities plus the velocities, or using an interpolated line going through the overdensities plus the velocities (see Sec.~\ref{sec:results} for a detailed description). The accuracy is defined as (median MCMC value)/(actual simulation value). The first 10 parameters are the actual modeling parameters for: the halo (Sec.~\ref{sssec:halo_sim}), the cluster  (Sec.~\ref{sssec:pal5_sim}), and the Sun (Sec.~\ref{sssec:sun_sim}), whereas the last 5 parameters are derived quantities (see Tab.~\ref{tab:accuracies}).}
\label{sim_summary}
\end{figure*}

\begin{figure*}
\centering
\includegraphics[width=0.9\textwidth]{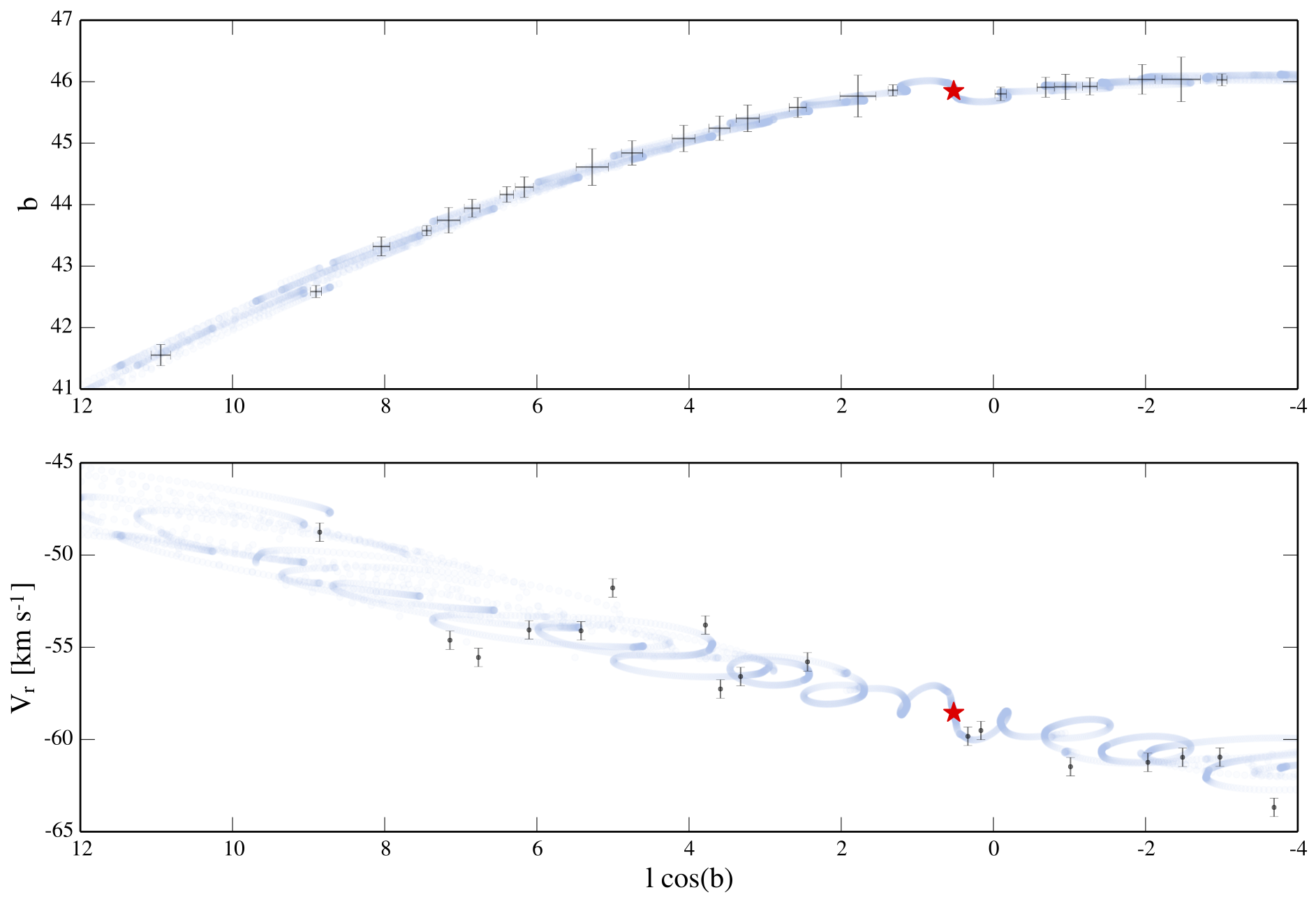}
\caption{\textit{Streakline} model (blue semi-transparent points) of the Pal\,5-analog simulation using the best-fit parameters from the weighted-overdensity analysis approach. The present-day position of the cluster is marked by a red star, whereas as the locations and uncertainties of overdensities (upper panel) and radial velocity stars (lower panel) are shown by black markers.}
\label{winning_model_sim}
\end{figure*}

As described in Sec.~\ref{sssec:pal5analog}, we ran a direct $N$-body simulation of a star cluster on a Pal\,5-like orbit within a Milky-Way like potential to test our methodology and assess its accuracy. A summary of this assessment is shown in Fig.~\ref{sim_summary} and listed in Tab.~\ref{tab:accuracies}. The accuracy in this plot is defined as (median MCMC value)/(actual simulation value). The error bars represent the precision of the measurements, that is, the 68\% confidence intervals. We note that more than 2/3 of the values are recovered within these 68\% confidence intervals, which indicates that we may in fact be overestimating our uncertainties.\\\\
All four analysis approaches recover the 10 true simulation parameters (and also the 5 derived quantities shown in Fig.~\ref{sim_summary}) within their respective uncertainties -- with only few exceptions, which we will discuss in detail in the following. This is very reassuring and shows that our methodology works.\\\\
Both the precision and the accuracy of recovering the true simulation values vary significantly between the model parameters as well as between the methods. In general, we find that using only the overdensities produces the largest uncertainties (lowest precision). This is not surprising since more data also means more constraining power. On average, the uncertainties get significantly smaller by adding the 17 radial velocity stars to our modeling. The exact gain in precision varies from parameter to parameter, but lies mostly between 10-30\%.\\\\
The precision can be further improved by adding more information, that is, by adding weights to the overdensities, or by interpolating between them. The run with weighted overdensities and the run with the interpolated centerline turn out to achieve similar results in terms of precision. The gain in precision compared to the run with the overdensities and radial velocities can be as large as 50\%. From Tab.~\ref{tab:accuracies} it is apparent that the weighted overdensity analysis approach achieves a significantly better accuracy in recovering the simulation parameters than the interpolated centerline analysis approach. The latter can even introduce such strong biases that the true model value lies outside the 68\% confidence interval.\\\\
In Fig.~\ref{winning_model_sim} we show the best-fit model from the weighted overdensity approach. The \textit{streakline} model perfectly reproduces the shape of the tails and the radial velocity gradient. The $N$-body simulation created a regular overdensity pattern, which originated from epicyclic motion within the tails. This pattern has been successfully recovered by our \textit{Difference-of-Gaussians} overdensity finder after we inserted the $N$-body model into the SDSS field. After the MCMC modeling of the detected overdensities, the best-fit \textit{streakline} model nicely reproduces this regular pattern in the vicinity of the cluster, and the stream centerline follows the detected overdensities further away from the cluster, where the epicyclic overdensities are expected to be washed out. \textit{This demonstrates that epicyclic overdensities can be used to constrain models of tidal streams.}\\\\
The results from the different analysis approaches will be discussed in detail, and separately for the different model-parameter groups, in the following.

\subsubsection{Halo parameters results (analog simulation)}\label{sssec:halo_sim}
\begin{figure}
\centering
\includegraphics[width=0.45\textwidth]{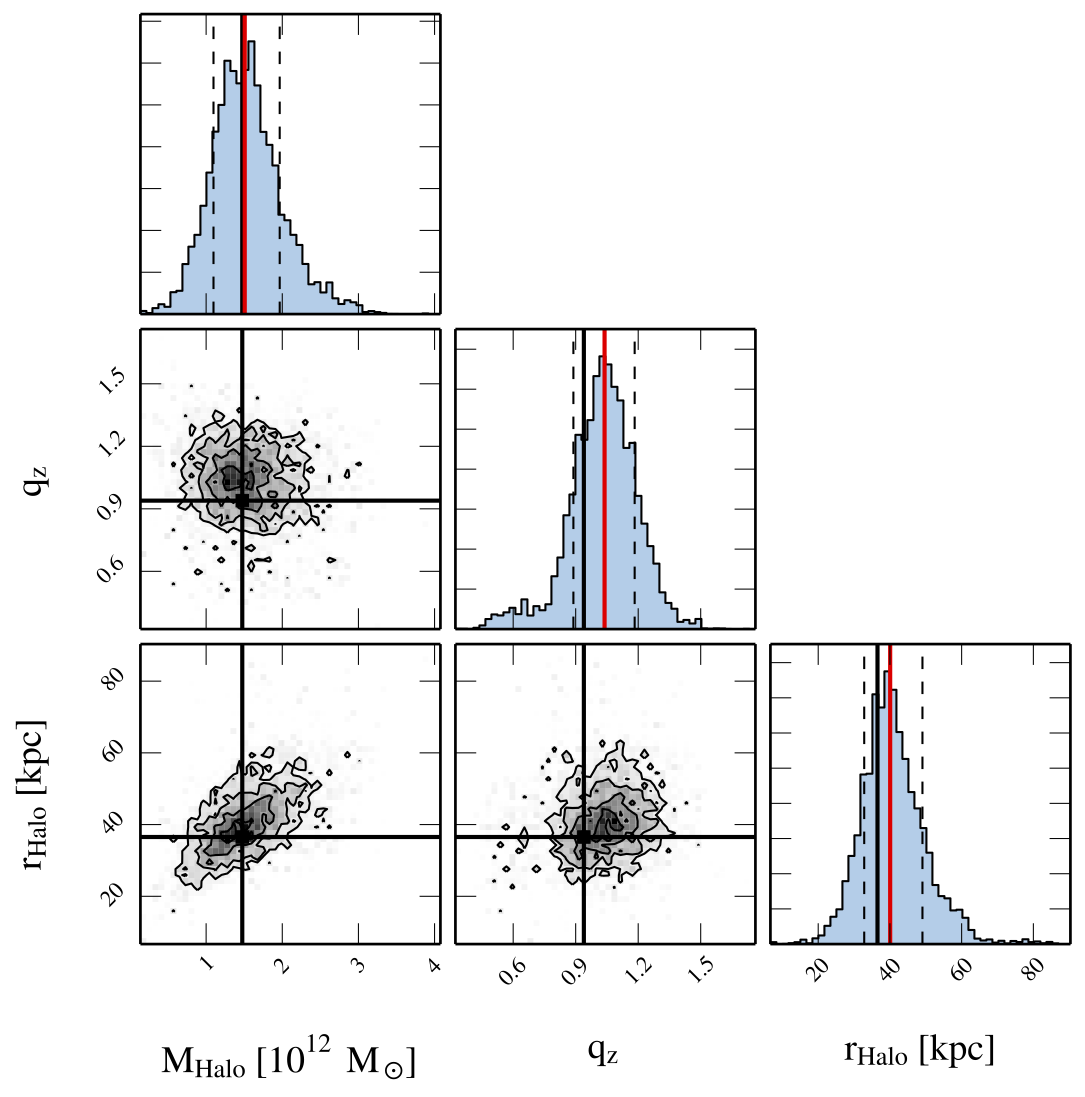}
\caption{Posterior parameter distributions of the three halo parameters (scale mass, $M_{Halo}$; scale radius, $r_{Halo}$; flattening along z-axis, $q_z$; see Eq.~\ref{eq:halo}). The blue-shaded histograms show the marginalized posterior distributions for each parameter, the red lines indicate the median values, the dashed lines give the 68\% confidence intervals, and the black vertical lines mark the true simulation values. The contour plots show the two-dimensional posterior distributions for parameter pairs. Uncorrelated pairs should show roundish contours, whereas correlated parameters like $r_{Halo}$ and $M_{Halo}$ have elongated distributions. All model parameters are recovered within the respective uncertainties.}
\label{sim_tri_halo}
\end{figure}

All four methods recover the three halo parameters, i.e.~$M_{Halo}$, $r_{Halo}$ and $q_z$,  of our analog simulation within the uncertainties. The left side of Fig.~\ref{sim_summary} shows the results for the halo parameters for the four analysis approaches. As can also be seen from Tab.~\ref{tab:accuracies}, all parameters are recovered within about 20\% in terms of accuracy by all methods. The best accuracy is achieved by the weighted overdensity analysis approach. The interpolated centerline method yields a comparable precision, but introduces biases which cause, e.g., the true halo flattening parameter to lie just outside the 68\% confidence interval.\\\\
The largest uncertainties (lowest precision) we get in the scale length of the halo when we only use the overdensities, with $1\,\sigma$ uncertainties of $+59\%$ and $-37\%$. This is expected since we chose the scale length of our NFW halo to be 36.5\,kpc, which is well outside the apogalactic radius of our Pal\,5-analog's orbit. Moreover, due to the parametrization of this halo, there is a significant degeneracy between the scale length and the scale mass of the halo. This can be seen in Fig.~\ref{sim_tri_halo}, where we show the posterior parameter distributions for the halo parameters for the analysis approach using the weighted overdensities. The three blue-shaded histograms give the marginalized posterior distributions for the three respective parameters. The median of this distribution is indicated by a red vertical line, and the 68\% confidence intervals are marked by dashed vertical lines. The true parameter value from the analog simulation is shown by a  black solid line. Ideally, the red line should lie on top of the  black.\\\\
According to Fig.~\ref{sim_summary} we can recover an underlying gravitational potential with a Pal\,5-like stream to an accuracy of $\approx10$\% and with an uncertainty of 20-30\% in each halo parameter when we use the weighted overdensity analysis approach. This also holds for quantities derived from our model parameters, keeping in mind that the parametrizations of the Galactic potential in the $N$-body simulation and in the \textit{streakline} modeling is identical. In Fig.~\ref{sim_summary}, we also show the (more frequently used) virial mass, $M_{200}$, virial radius, $r_{200}$, and concentration, $c$, derived from the scale masses and scale lengths of our trial models. As expected, the uncertainties on these quantities are comparable to the ones of the original model parameters.\\\\
Interestingly, the acceleration at the (assumed) location of Pal\,5, $a_{Pal\,5}$, shows a factor of 2 lower spread than, e.g., $M_{200}$ (Tab.~\ref{tab:accuracies}). This is due to the fact that scale mass and scale radius of the halo are strongly degenerate (see lower left panel of Fig.~\ref{sim_tri_halo}), and a wide range of combinations of these two quantities can produce very similar accelerations at the location of Pal\,5. \textit{Reporting accelerations at given points within the Milky Way halo, instead of halo masses within a certain radius, or virial masses, may therefore be a more reasonable way of comparing results from different studies - especially if the halo is found to be aspherical, since then the term 'enclosed mass' is ill-defined.}

\subsubsection{Palomar\,5 parameter results (analog simulation)}\label{sssec:pal5_sim}
\begin{figure*}
\centering
\includegraphics[width=0.71\textwidth]{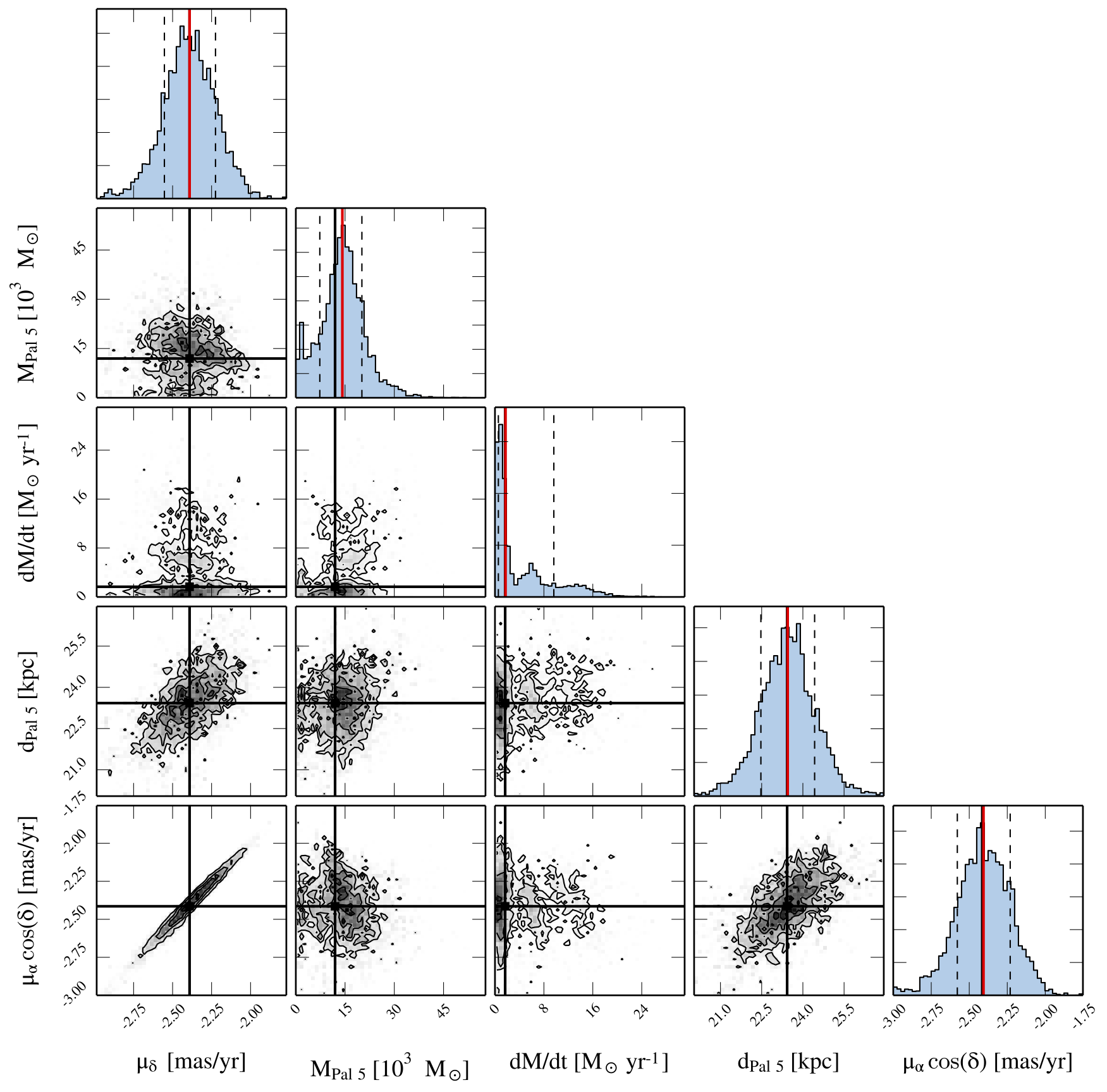}
\caption{Same as Fig.~\ref{sim_tri_halo} but for the cluster parameters (present-day mass of Pal\,5, $M_{Pal\,5}$; mass loss rate, $dM/dt$; heliocentric distance, $d_{Pal\,5}$; proper motion components, $\mu_\alpha\cos(\delta)$ and $\mu_\delta$). All model parameters from the analog simulation (black lines) are recovered within the 68\% uncertainty intervals (dashed lines).}
\label{sim_tri_cluster}
\end{figure*}

For our \textit{streakline} models, five out of 10 model parameters determine the properties of each trial cluster. The heliocentric distance of Pal\,5, $d_{Pal\,5}$, and its two proper motion components, $\mu_\alpha\cos(\delta)$ and $\mu_\delta$, set its orbit. The present-day mass, $M_{Pal\,5}$, and mass loss rate, $dM/dt$, fix its mass evolution. All of these cluster parameters are recovered within the 68\% confidence intervals from all four methods (Tab.~\ref{tab:accuracies}) -- the only exception is again the interpolated centerline method, which fails at recovering the cluster mass.\\\\
From Fig.~\ref{sim_summary} it is obvious that the three orbital parameters are significantly better constrained through the data than the mass evolution parameters. The weaker performance on these mass parameters is most certainly due to the weak dependence of the tidal radius (alas the key quantity that links the tails and the cluster) on the cluster mass (see Eq.~\ref{eq:tidalradius}). Nevertheless, all analysis approaches except for the interpolated centerline method recover the cluster mass, even though the departures can be as large as $+123\%$ and $-83\%$ if we use just the overdensities. \textit{Yet, with a precision of about $\pm 50\%$, the weighted overdensities analysis approach provides a robust measure for a quantity that is hard to determine observationally} (see Sec.~\ref{ssec:discussion}).\\\\
The mass loss rate is only weakly constrained and shows an asymmetric posterior distribution with a tail towards high mass-loss rates (Fig.~\ref{sim_tri_cluster}). Even so, it is intriguing that the shape of a tidal stream can give insights on the mass evolution of its progenitor.\\\\
The full strength of tidal streams as high-precision scales can be seen in the orbital parameters. The precision and accuracy of the orbital parameters is better than $\approx10\%$ for all four methods. The weighted overdensity analysis approach recovers the distance to the mock Pal\,5 with an accuracy of 1.00 and a precision of $\pm4\%$. The proper motion components are also accurately recovered by this approach to a precision of $\approx7\%$. As can be seen in Fig.~\ref{sim_tri_cluster}, the two components are highly correlated due to relatively close alignment of the tails and the cluster orbit, that is, the direction of the cluster's transverse direction of motion is very well constrained through the tails, whereas the cluster's transverse speed allows for some variation.

\subsubsection{Solar parameter results (analog simulation)}\label{sssec:sun_sim}
\begin{figure}
\centering
\includegraphics[width=0.325\textwidth]{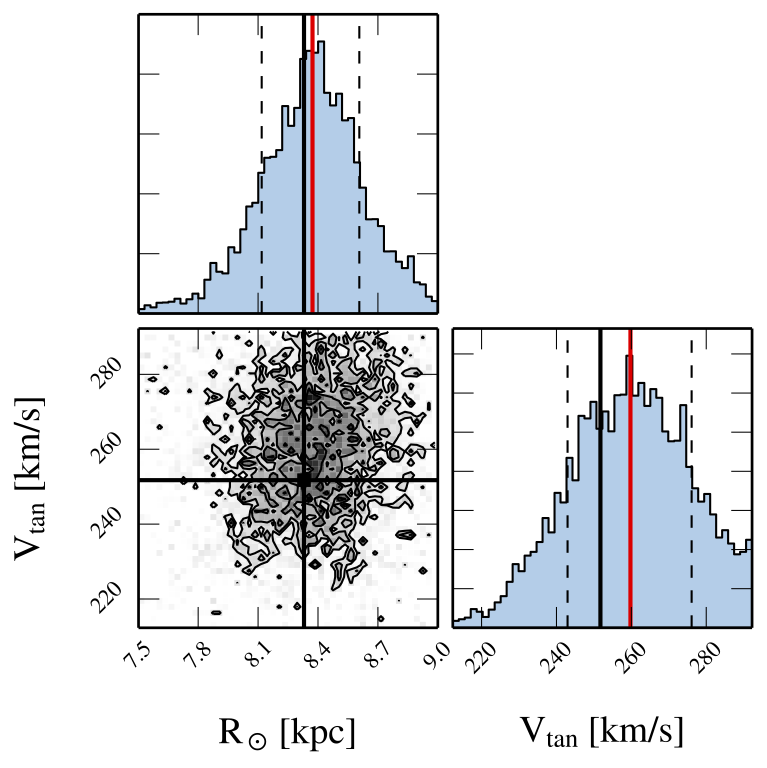}
\caption{Same as Fig.~\ref{sim_tri_halo} but for the Solar parameters (Solar transverse velocity with respect to the Galactic center, $V_{tan}$; Galactocentric radius, $R_{\odot}$). The Solar parameters from the analog simulation (black lines) are recovered with high accuracy.}
\label{sim_tri_orbital}
\end{figure}

Just like for Pal\,5's distance and its proper motion, the two Solar parameters in our modeling, $V_{tan}$ and $R_\odot$, are recovered with remarkable accuracy by all four methods. Using the weighted overdensities, the precision can be as good as $7\%$ for the Solar transverse velocity, and $3\%$ for the Galactocentric radius of the Sun. \textit{This clearly demonstrates the power of cold tidal streams in constraining length and velocity scales}\footnote{As long as they lie within the orbit of the progenitor.}.\\\\
As a sanity check, we computed the circular velocities at the assumed Galactocentric radius of the Sun, $V_C(R_\odot)$, for all parameter sets in our MCMC chains. All methods recover the expected value with high accuracy and precision (Tab.~\ref{tab:accuracies}). This is reassuring, but should be taken with a grain of salt as in both $N$-body simulation and \textit{streakline} modeling we used the same (fixed) potentials for the disk and the bulge (see Sec.~\ref{ssec:potential}). Moreover, in both cases we used the same parametrization for the dark halo potential, which for the real Milky Way we unfortunately do not know. This should be kept in mind when interpreting the results from the observations of Pal\,5.

\subsection{Palomar 5 observations}\label{ssec:resultspal5}

\begin{table*}
\label{tab:results}
\centering
\caption{Results from fitting to the observations}
\begin{tabular}{l|cccc}
 & Overdensities & Overdensities & Weighted & Interpolated\\
Parameter                  &                       &  + radial velocities &   overdensities  & centerline\\ \hline
$M_{Halo}\,[10^{12}\msun ]$ & $1.60^{+0.72}_{-0.63}$  & $1.75^{+0.76}_{-0.66}$   & $1.58^{+0.38}_{-0.37}$  & $1.81^{+0.45}_{-0.43}$   \\
$r_{Halo}$\,[kpc] & $33.2^{+21.5}_{-13.6}$  & $41.8^{+14.5}_{-11.0}$   & $37.9^{+9.1}_{-7.8}$  & $37.0^{+8.4}_{-6.8}$   \\
$q_{z}$ & $1.01^{+0.23}_{-0.24}$  & $0.84^{+0.27}_{-0.16}$   & $0.95^{+0.16}_{-0.12}$  & $1.00^{+0.16}_{-0.12}$   \\\hline
$M_{Pal\,5}\,[10^3\msun]$ & $16.8^{+15.3}_{-11.2}$  & $18.8^{+12.9}_{-10.6}$   & $16.0^{+8.5}_{-5.9}$  & $14.5^{+9.9}_{-7.8}$   \\
$dM/dt\,[10^3\msun\,\mbox{Gyr}^{-1}]$ & $4.6^{+6.7}_{-3.5}$  & $10.1^{+6.5}_{-8.1}$   & $7.9^{+5.3}_{-6.4}$  & $1.9^{+2.4}_{-1.0}$   \\
$d_{Pal\,5}$\,[kpc] & $23.52^{+1.23}_{-1.17}$  & $23.50^{+1.17}_{-1.10}$   & $23.58^{+0.84}_{-0.72}$  & $23.12^{+0.87}_{-1.00}$   \\
$\mu_{\alpha}\cos(\delta)$\,[mas\,yr$^{-1}$] & $-2.49^{+0.25}_{-0.28}$  & $-2.40^{+0.21}_{-0.23}$   & $-2.39^{+0.15}_{-0.17}$  & $-2.50^{+0.17}_{-0.18}$   \\
$\mu_{\delta}$\,[mas\,yr$^{-1}$] & $-2.43^{+0.22}_{-0.24}$  & $-2.38^{+0.20}_{-0.21}$   & $-2.36^{+0.14}_{-0.15}$  & $-2.44^{+0.15}_{-0.16}$   \\\hline
$V_{tan}$\,[km\,s$^{-1}$] & $252.9^{+26.2}_{-23.8}$  & $254.5^{+20.6}_{-21.4}$   & $254.3^{+15.7}_{-16.1}$  & $251.1^{+18.4}_{-17.0}$   \\
$R_{\odot}$\,[kpc]  & $8.32^{+0.37}_{-0.42}$  & $8.35^{+0.33}_{-0.35}$   & $8.30^{+0.24}_{-0.25}$  & $8.29^{+0.30}_{-0.31}$   \\\hline
$V_{C}(R_\odot)$\,[km\,s$^{-1}$] & $241.7^{+41.8}_{-24.6}$  & $230.6^{+15.5}_{-12.8}$   & $233.0^{+12.7}_{-10.0}$  & $238.2^{+11.5}_{-9.3}$   \\
$a_{Pal\,5}$\,[pc\,Myr$^{-2}$] & $2.91^{+1.65}_{-0.99}$  & $2.60^{+0.87}_{-0.63}$   & $2.61^{+0.56}_{-0.44}$  & $2.89^{+0.53}_{-0.48}$   \\
$M_{200}\,[10^{12}\msun]$ & $1.70^{+1.17}_{-0.84}$  & $1.59^{+0.80}_{-0.68}$   & $1.56^{+0.42}_{-0.42}$  & $1.85^{+0.49}_{-0.43}$   \\
$r_{200}$\,[kpc] & $200.5^{+38.2}_{-41.0}$  & $196.0^{+28.5}_{-33.5}$   & $194.8^{+16.3}_{-19.2}$  & $206.3^{+16.9}_{-17.6}$   \\
$c$ & $6.15^{+4.87}_{-2.95}$  & $4.74^{+1.73}_{-1.49}$   & $5.14^{+1.54}_{-1.16}$  & $5.56^{+1.33}_{-1.06}$   \\
\end{tabular}
\end{table*}

\begin{figure*}
\centering
\includegraphics[width=0.9\textwidth]{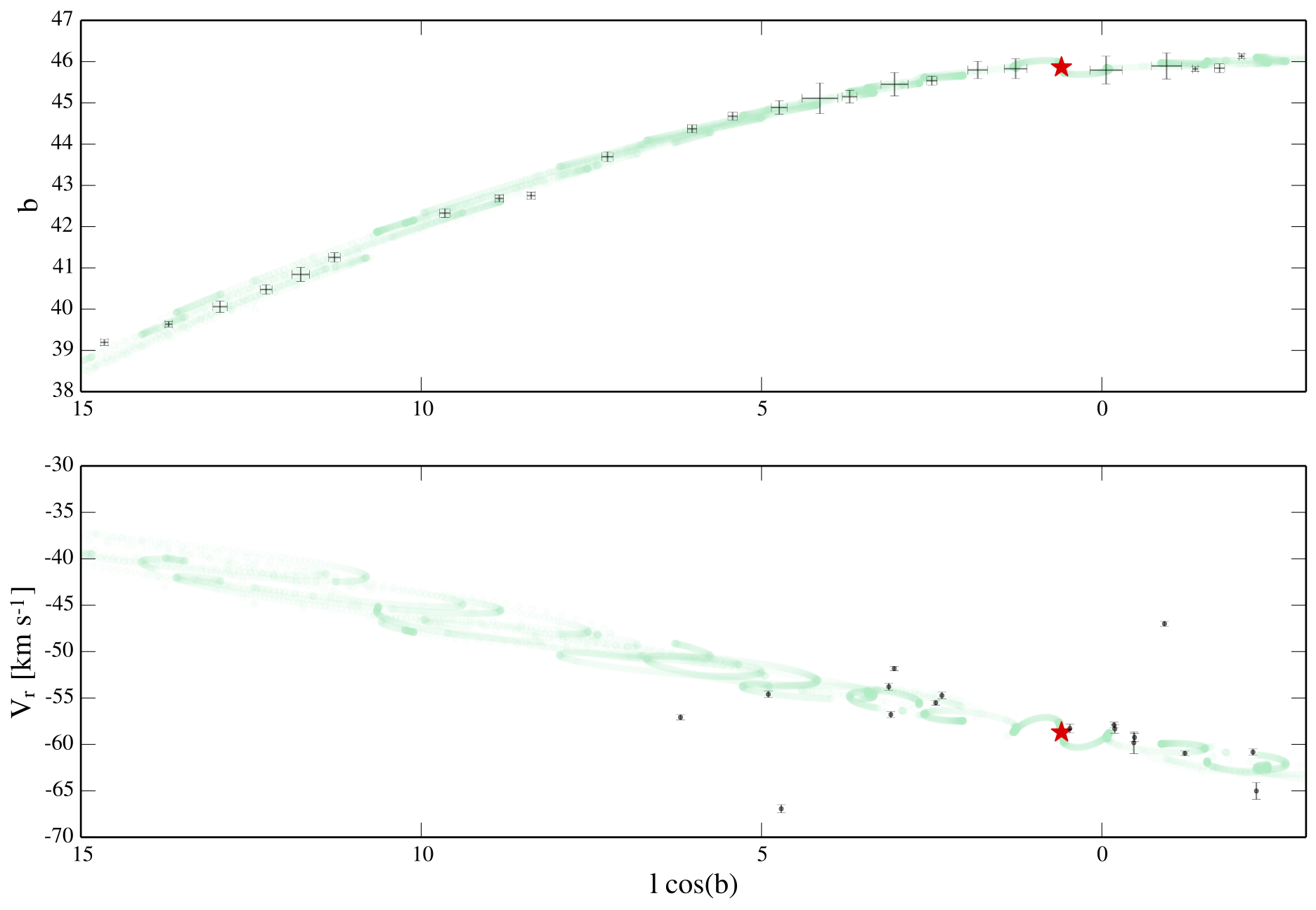}
\caption{\textit{Streakline} model (green semi-transparent points) of Pal\,5 using the best-fit parameters from the weighted-overdensity analysis approach. The present-day position of the cluster is marked by a red star, whereas the locations and uncertainties of overdensities (upper panel) and radial velocity stars (lower panel) are shown by black markers. Just like for the Pal\,5-analog simulation, the \textit{streakline} model traces the periodic overdensities very accurately close to the cluster.}
\label{winning_model}
\end{figure*}

The posterior parameter distributions of our modeling of the Pal\,5 observations look basically the same as the ones from the analog simulation (see Fig.~\ref{obs_tri_halo}-\ref{obs_tri_orbital}), just with slightly different median and uncertainty values (Tab.~\ref{tab:results}). This was to be expected since we chose the parameters of our analog $N$-body simulation based on preliminary modeling of the observations. Hence, since the analog simulation probes a configuration that is very similar to the real Pal\,5, the underlying systematic uncertainties should also be comparable.\\\\
For the weighted-overdensity approach we show the median model in Fig.~\ref{winning_model}. The epicylcic loops of the stream stars are clearly visible and the closest loops can be associated with the overdensities we found in the SDSS data. At about 5\,deg distance from the cluster, the epicyclic loops overlap so much that no clear substructure is generated, especially since our \textit{streakline} models have no scatter in their escape conditions, and hence in a more realistic simulation the structures would wash out more easily.\\\\
We will discuss the results from the different analysis approaches in detail for the three parameter groups in the following.

\subsubsection{Halo parameter results (observations)}\label{sssec:halo_obs}

\begin{figure}
\centering
\includegraphics[width=0.45\textwidth]{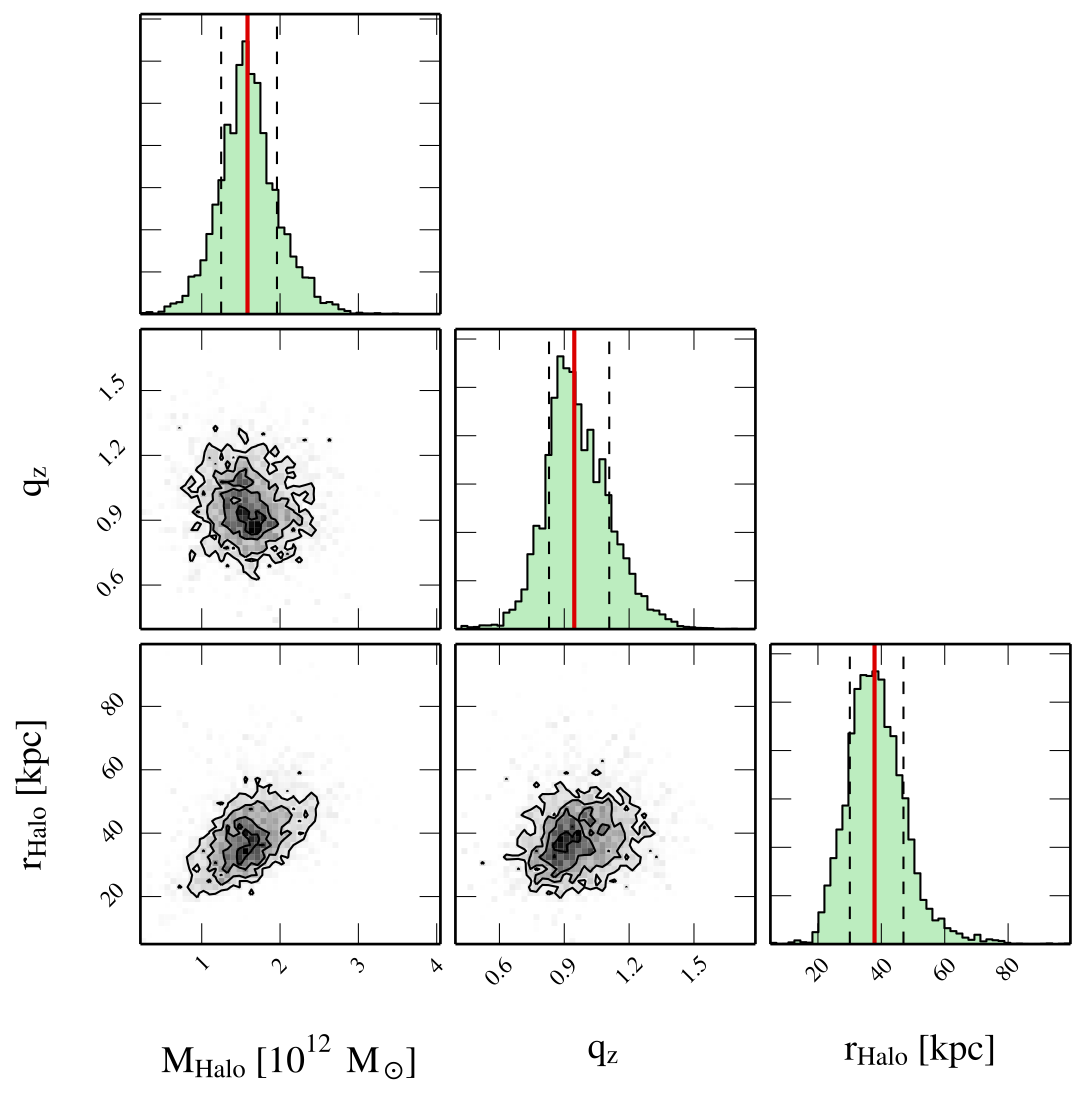}
\caption{Posterior parameter distributions of the three halo parameters (scale mass, $M_{Halo}$; scale radius, $r_{Halo}$; flattening along z-axis, $q_z$; see Eq.~\ref{eq:halo}) as recovered from the Pal\,5 observations using the weighted overdensity analysis approach. The green-shaded histograms show the marginalized posterior distributions for each parameter, the red lines indicate the median values, and the dashed lines give the 68\% confidence intervals. Our best-fit NFW halo is slightly oblate, has a large scale radius, and a mass of about $(1.6\pm0.4)\times 10^{12}\msun$.}
\label{obs_tri_halo}
\end{figure}

Just like for the analog simulation, we get similar results from all four analysis approaches, and similar to the analog simulation results, the results from the observations also get more precise the more constraints we use (Tab.~\ref{tab:results}). Moreover, all results for the halo parameters from the different approaches agree with each other within their respective 68\% confidence intervals. In the following we will discuss the results from the approach using the weighted overdensities as it shows, on average, the smallest uncertainties. The results from this analysis approach are very similar to the results using the interpolated centerline.\\\\
Assuming that the dark halo potential of the Milky Way has an NFW shape, we get from our modeling of the Pal\,5 observations that the halo is best represented by a scale mass of $M_{Halo} = (1.6\pm0.4)\times 10^{12}\msun$, a scale radius of $r_{Halo} = 38^{+9}_{-8}$\,kpc, and a flattening along the z-axis of $0.95^{+0.16}_{-0.12}$ (Fig.~\ref{obs_tri_halo}). The respective virial mass, virial radius and concentration are $M_{200} = (1.6\pm0.4)\times 10^{12}\msun$, $r_{200} = 195^{+16}_{-19}$\,kpc, and $c = 5.1^{+1.5}_{-1.2}$. This is higher than most recent estimates of the Milky Way's virial mass, and also the concentration value is significantly lower than most studies find (see Sec.~\ref{ssec:discussionhaloparameters}). This may be due to the fact that the scale radius of the assumed NFW halo lies well outside Pal\,5's orbit (apogalactic radius of $\approx19$\,kpc) and is therefore a vast  extrapolation out to about 200\,kpc, and/or that the NFW potential is a poor representation of the Galactic potential in the inner 19\,kpc, and/or that we fixed the disk and bulge potentials.\\\\
A more sensible and also better constrained quantity is the acceleration at the present-day position of Pal\,5. We find that $a_{Pal\,5} = 2.6^{+0.6}_{-0.4}$\,pc\,Myr$^{-2}$, which corresponds to $(0.81^{+0.17}_{-0.14})\times 10^{-10}$\,m\,s$^{-2}$ (Tab.~\ref{tab:results}). Using Eq.~\ref{eq:vcpal5}, this corresponds to a local circular velocity of $V_C\left(r_{Pal\,5}\right) = 217^{+22}_{-19}$\,km\,s$^{-1}$ at the Galactocentric radius of Pal\,5 ($r_{Pal\,5}=18.8^{+0.8}_{-0.7}$\,kpc). From the acceleration we can also calculate an ``equivalent mass'', that is, the enclosed mass assuming the mass inside the present-day radius of Pal\,5 was spherically distributed:
\begin{equation}
a_{Pal\,5}\,r_{Pal\,5}^2\,G^{-1} = (2.06^{+0.44}_{-0.37})\times10^{11}\msun.
\end{equation}
In the symmetry plane of our Galactic potential we do not have these problems of quoting sensible values for comparison with other studies. A common way to express the acceleration or enclosed mass within the galactic disk is the circular velocity. Even though we seem to limit ourselves by choosing a fixed parametrization for the dark halo, we find that all analysis approaches produce median posterior values for the circular velocity at the Solar radius, $V_C(R_\odot)$, between 231\,km\,s$^{-1}$ and 242\,km\,s$^{-1}$. That is, all recovered halo parameters produce reasonable halos that are in agreement with most estimates for the circular velocity at the Solar circle (see Sec.~\ref{ssec:discussion} for a more detailed comparison to other observational constraints). This is reassuring and tells us that our choice of parametrization is at least not obviously wrong. The weighted overdensity approach yields a circular velocity of $V_C(R_\odot) = 233.0^{+12.7}_{-10.0}$\,km\,s$^{-1}$.\\\\
As we will see in the following, our results for the other model parameters also agree with independent estimates from the literature.

\subsubsection{Palomar\,5 parameter results (observations)}\label{sssec:pal5_obs}

\begin{figure*}
\centering
\includegraphics[width=0.71\textwidth]{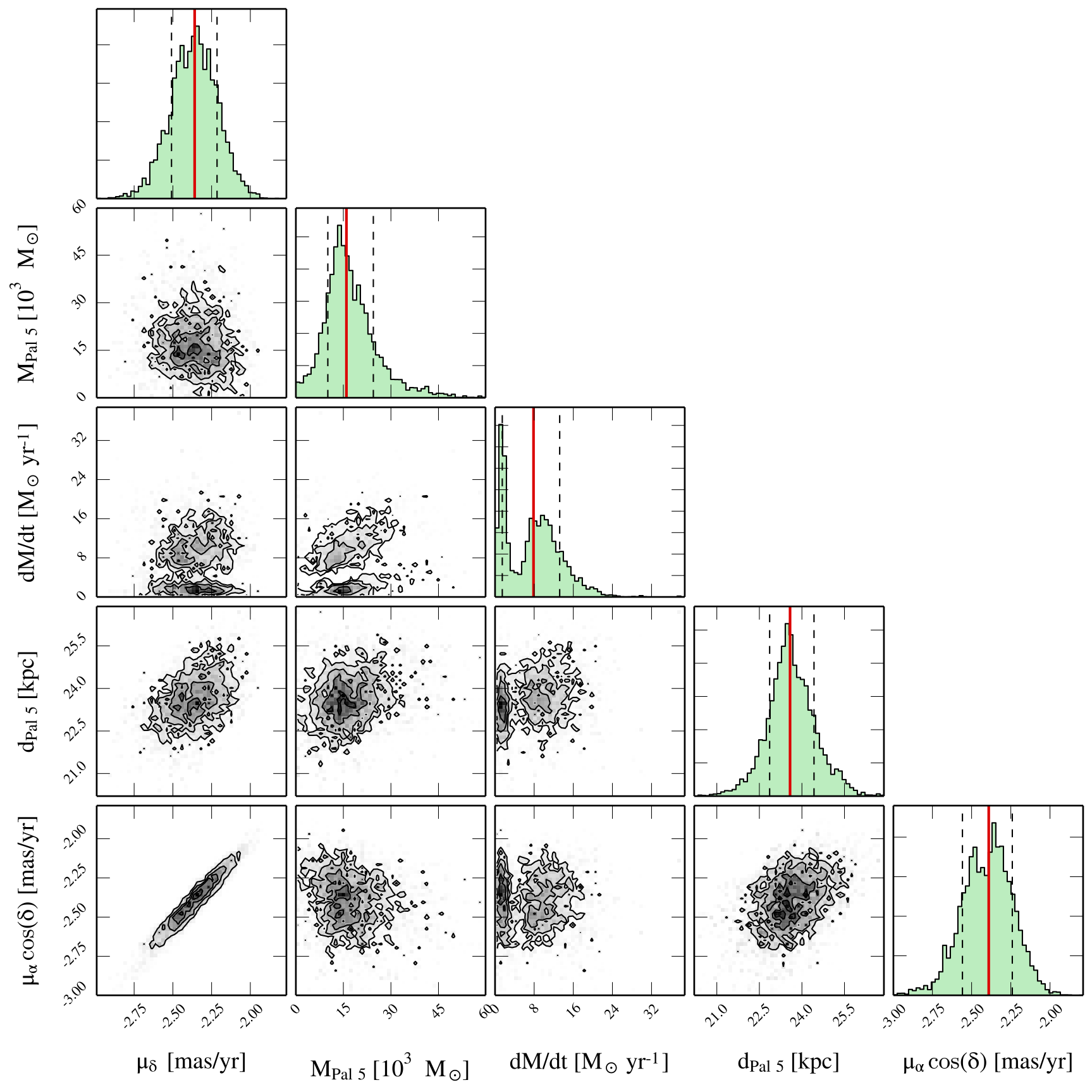}
\caption{Same as Fig.~\ref{obs_tri_halo} but for the cluster parameters (present-day mass of Pal\,5, $M_{Pal\,5}$; mass loss rate, $dM/dt$; heliocentric distance, $d_{Pal\,5}$; proper motion components, $\mu_\alpha\cos(\delta)$ and $\mu_\delta$).}
\label{obs_tri_cluster}
\end{figure*}

\begin{table*}
\label{tab:orbits}
\centering
\caption{Orbital parameters of the best-fit model of each analysis approach}
\begin{tabular}{l|cccc}
  &  Overdensities & Overdensities + radial velocities & Weighted overdensities  & Interpolated centerline\\
 \hline
Present-day distance & 18.70 kpc & 18.68 kpc & 18.77 kpc & 18.34 kpc \\
Apogalactic distance & 19.06 kpc  & 19.07 kpc & 19.15 kpc  & 18.67 kpc \\
Perigalactic distance & 8.30 kpc & 7.83 kpc & 7.54 kpc & 7.97 kpc \\
Orbital eccentricity & 0.39  & 0.42  & 0.43 & 0.40 \\
\end{tabular} 
\end{table*}

The posterior parameter distributions for the 5 cluster parameters (Fig.~\ref{obs_tri_cluster}) look remarkably similar in extent and shape to the posterior distributions from the analog simulation (cf.~Fig.~\ref{sim_tri_cluster}). With the weighted overdensity analysis approach, we infer a cluster mass of $M_{Pal\,5} = (16.0^{+8.5}_{-5.9})\times 10^3\msun$.\\\\
As with nearly all other parameters, the four approaches produce similar results, which all agree within their 68\% confidence intervals with each other (Tab.~\ref{tab:results}). One exception is the mass loss rate, for which the interpolated centerline method yields a comparatively low value with small uncertainties. We haven't observed such a bias in this particular parameter for the analog simulation, hence we cannot relate this discrepancy to any known effect. Looking at the posterior distribution of this generally weakly constrained parameter, however, we find that it tends to be asymmetric and bimodal (Fig.~\ref{obs_tri_cluster}). We already pointed this out for the Pal\,5 analog, but here the bimodality appears to be more pronounced. The interpolated centerline approach seems to strongly prefer the peak with the lower mass loss rate. The other analysis approaches yield larger values for $dM/dt$ but also larger uncertainties.\\\\
The proper motion in right ascension shows a slightly broadened distribution with respect to the analog simulation, even though the same parameters also showed a lower kurtosis for the mock Pal\,5. For the observed Pal\,5 the posterior distribution in $\mu_\alpha\cos(\delta)$ may even be bimodal. Since $\mu_\alpha\cos(\delta)$ and $\mu_\delta$ are strongly correlated, the proper motion in declination also shows a broader distribution than we got for the analog simulation. The median proper motion values and the corresponding 68\% confidence intervals are $\mu_\alpha\cos(\delta) = -2.39^{+0.15}_{-0.17}$\,mas\,yr$^{-1}$ and $\mu_\delta = -2.36^{+0.14}_{-0.15}$\,mas\,yr$^{-1}$, which corresponds to a precision of better than $7\%$.\\\\
The other two cluster parameters are well behaved.  Pal\,5's heliocentric distance we recover with a remarkable precision of $3\%$ to be $23.6^{+0.8}_{-0.7}$\,kpc. As for the mock Pal\,5, we find that the geometry of Pal\,5's tidal stream is very powerful in constraining length and velocity scales.\\\\
 It is reassuring that the four different analysis approaches for comparing the \textit{streakline} models to the observational data points yield comparable results. The models using the median values of the posterior parameter distributions from the four approaches are hardly distinguishable from each other, and also the orbital parameters of the median models all fall within a narrow range (Tab.~\ref{tab:orbits}).

\subsubsection{Solar parameter results (observations)}\label{sssec:sun_obs}

\begin{figure}
\centering
\includegraphics[width=0.325\textwidth]{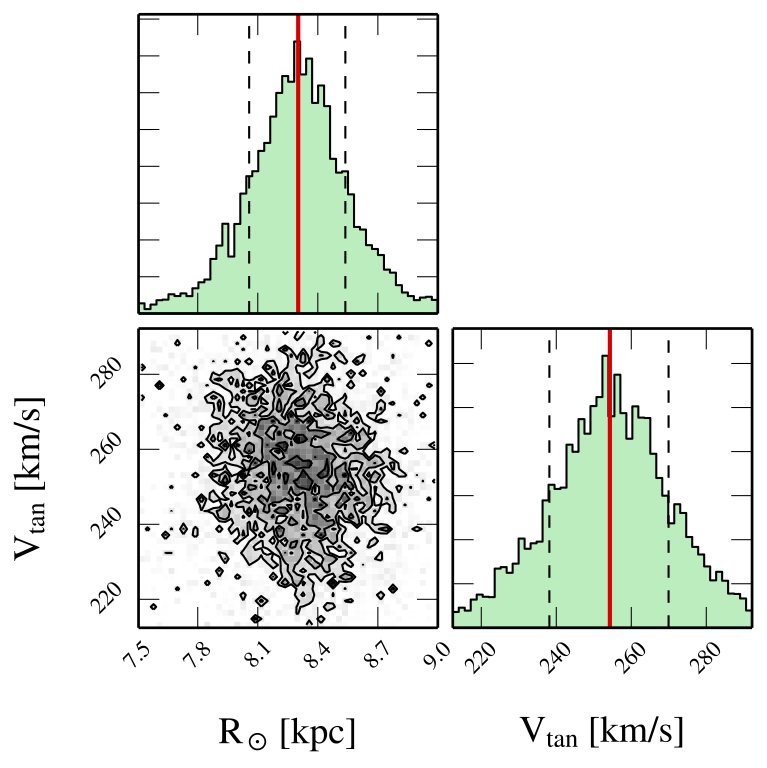}
\caption{Same as Fig.~\ref{obs_tri_halo} but for the Solar parameters (Solar transverse velocity with respect to the Galactic center, $V_{tan}$; Galactocentric radius, $R_{\odot}$). }
\label{obs_tri_orbital}
\end{figure}

As expected from the modeling of the analog simulation, the two parameters of our \textit{streakline} models that describe the location and velocity of the Sun with respect to the Galactic rest frame, and hence to Pal\,5, turn out to be very well constrained. For the transverse velocity of the Sun with respect to the Galactic center we get $V_{tan} = 254\pm16$\,km\,s$^{-1}$, and for the Galactocentric distance of the Sun we get $R_\odot = 8.30^{+0.24}_{-0.25}$\,kpc. Again, the four analysis approaches yield statistically indistinguishable results. Moreover, these values (and others of the 10 model parameters) agree very well with independent determinations from the literature. These will be discussed in the following.

\subsubsection{Hyper-parameter results (observations)}\label{sssec:hyper}

In Sec.~\ref{ssec:comparison} we described the four hyper-parameters of our modeling process, that is, the integration time, the smoothing in positional uncertainty of the overdensities, the smoothing in the positional uncertainties of the radial velocity stars, and the smoothing in the radial velocities. These were allowed to vary between zero and nearly arbitrarily large values. However, the posterior distributions for these hyper-parameters turn out to be absolutely reasonable and well-behaved. The distribution of integration times peaks at $3.4^{+0.5}_{-0.3}$\,Gyr, telling us that the parts of the Pal\,5 stream for which we detect significant overdensities within the SDSS footprint is of about this age. The positional uncertainty of the overdensities peaks at $(0.13\pm0.03)$\,deg, whereas the positional uncertainty of the radial velocity stars lies at $0.2^{+0.3}_{-0.2}$\,deg. Finally, the smoothing in radial velocity is $(0.7\pm0.3)$\,km\,s$^{-1}$. 

\section{Discussion}\label{ssec:discussion}
As we have shown in the previous Section, the four different approaches for comparing the \textit{streakline} models to the observational data yield results that are generally consistent among each other. This gives us confidence in the stability of the results, that is, the individual posterior parameter distributions for each model parameter are not artifacts of a single fitting method. To ensure that the analysis approaches do not fail collectively, we generated a simulation of a Pal\,5-analog on a similar orbit and in a similar potential as the real Pal\,5. Our methodology passed this test in that it successfully recovered all model parameters. Yet, the analog $N$-body simulation is quite idealized compared to the real Pal\,5: in the simulation we know for example the exact shape of the underlying galactic potential, and we fixed the mass and extent of the disk and bulge to some reasonable values. In order to test if the outcomes of the Pal\,5 modeling is reasonable and close enough to reality, we have to cross-check our results with estimates from other, independent methods. This is what we will do in the following for the three model parameter groups.

\subsection{Comparison of halo parameters to literature values}\label{ssec:discussionhaloparameters}

\begin{table}
\label{tab:halomass}
\centering
\caption{Mass of the Milky Way based on Pal\,5}
\begin{tabular}{l|l}
Measure & ``Weighted overdensities'' estimate\\
\hline
$M_{200}$ & $(1.69\pm{0.42})\times10^{12}\msun$\\
$M(r<100\,\mbox{kpc})$ & $(0.90\pm{0.20})\times10^{12}\msun$\\
$M(r<50\,\mbox{kpc})$ & $(0.43\pm{0.11})\times10^{12}\msun$\\
$M(r<r_{Pal\,5})_{disk}$ & $2.14^{+0.38}_{-0.35}\times10^{11}\msun$\\
$a_{Pal\,5}$ & $2.61^{+0.56}_{-0.44}$\,pc\,Myr$^{-2}$\\
$V_C(r_{Pal\,5})$ & $217^{+22}_{-19}$\,km\,s$^{-1}$\\
$a_{Pal\,5}\,r_{Pal\,5}^2\,G^{-1}$ & $2.06^{+0.44}_{-0.37}\times10^{11}\msun$
\end{tabular}
\end{table}

\begin{figure*}
\centering
\includegraphics[width=0.95\textwidth]{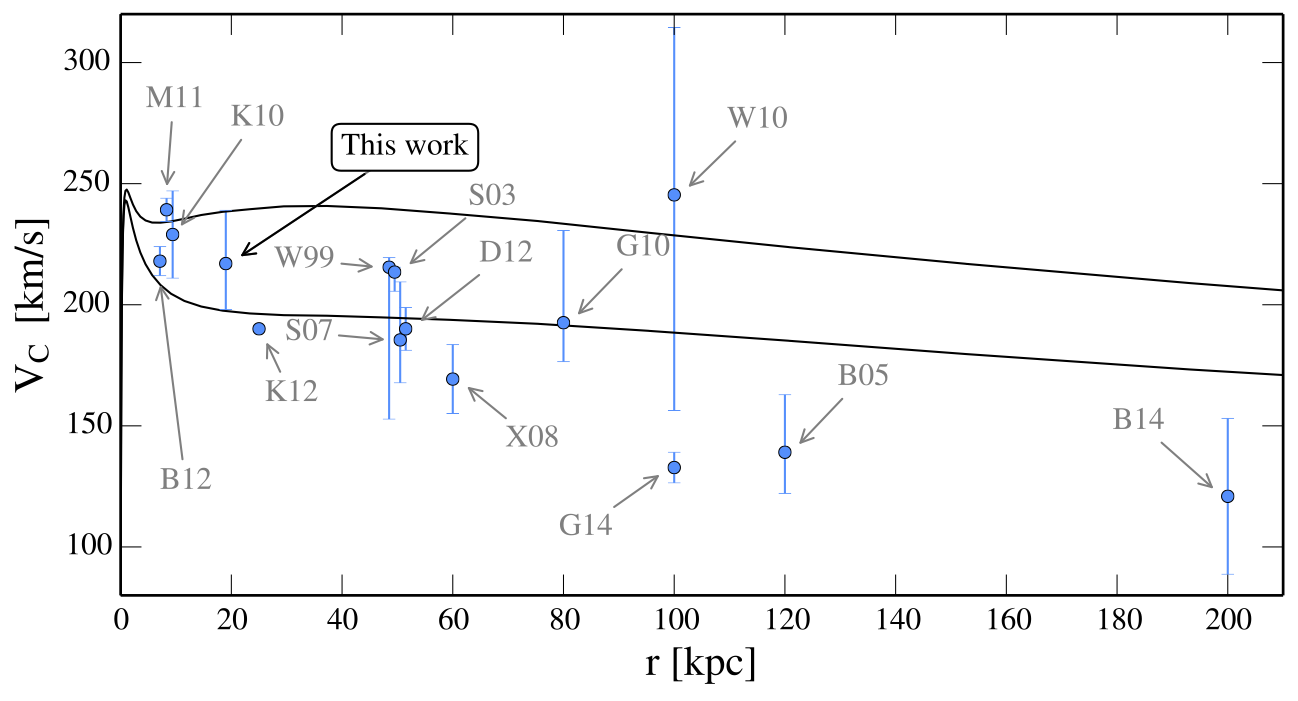}
\caption{Collection of circular velocity estimates from the literature. Our estimate for $R\approx19$\,kpc bridges the gap between disk tracer estimates and halo tracer estimates. The  black solid lines give the 68\% confidence intervals for the rotation curve resulting from our modeling of Pal\,5. The NFW parametrization that we chose for the dark halo potential does not seem to fall off quickly enough at radii larger than 19\,kpc to reproduce results from other investigations. References: W99 \citep{Wilkinson99}; S03 \citep{Sakamoto03}; B05 \citep{Battaglia05}; S07 \citep{Smith07}; X08 \citep{Xue08}; G10 \citep{Gnedin10}; K10 \citep{Koposov10}; W10 \citep{Watkins10}; M11 \citep{McMillan11}; B12 \citep{Bovy12}; D12 \citep{Deason12}; K12 \citep{Kafle12};  B14 \citep{Bhattacharjee14}; G14 \citep{Gibbons14}.}
\label{rotationcurve}
\end{figure*}

Weighing the Milky Way and its components has been one of the major goals in Galactic dynamics within the last few decades. Since the total mass of the Galaxy is vastly dominated by the extended dark halo, weighing the entire Galaxy basically boils down to determining the mass profile or gravitational potential of this halo. Various attempts have been made at constraining its virial mass, or the halo mass enclosed within a certain limiting radius, using all kinds of available tracers of the Galactic matter distribution. Some approaches also attempted to constrain the shape of the dark matter distribution, i.e.~its flattening or triaxiality. Here we list a few of the more recent measurements, first for the enclosed mass and then for the shape of the dark halo. 

\subsubsection{Enclosed mass}
Most of the estimates listed in the following are summarized as a rotation curve in Fig.~\ref{rotationcurve}, in which we collected enclosed mass estimates and circular velocity estimates from these references. However, we note that a comparison between the various studies is difficult due to the different measures quoted by the authors. Moreover, different tracers probe different radii, and some probe the Galactic potential in the MW disk, whereas others probe it at different points within the (not necessarily symmetric) halo. In an attempt to facilitate comparison, we compiled a list of ``classical'' measures from our own modeling of Pal\,5 (Tab.~\ref{tab:halomass}), but would like to point out that Pal\,5 essentially probes the Galactic mass within its apogalactic radius of $\approx 19$\,kpc. Our estimate for the MW's mass beyond that radius or its virial mass solely relies on the assumption that the dark halo potential is of NFW form. From our modeling there is no indication that this is the correct parametrization for the dark halo potential. In fact, most of the most recent studies indicate that the virial mass of the Milky Way halo is lower than our best-fit value. This may indicate that the NFW potential parametrization is a poor choice in the case of the Milky Way.\\\\
\citet{Irrgang13} fit three different, simple analytic galaxy models, each consisting of a disk, a bulge and a (spherical) halo, to a range of dynamical constraints. They compute various quantities for their models, of which we mention here only the mass enclosed inside a radius of 50 kpc, as it involves the least extrapolation. Depending on the parametrization they choose, they get values for $M(r<50\,\mbox{kpc})$ between $(0.46\pm 0.03)\times10^{12}\msun$ and $(0.81^{+0.13}_{-0.15})\times10^{12}\msun$. It is noteworthy that these estimates exclude each other, and solely differ in the choice of parametrization. We therefore focus on our mass estimate for Pal\,5's apogalactic radius of about 19\,kpc, and only give the other mass estimates for completeness.\\\\ 
However, since Pal\,5 is located well above the disk, it is outside any symmetry plane, which makes a definition of an enclosed mass difficult. A better estimate can be made, if we instead calculate the acceleration vector on Pal\,5 at its current position from the analytic potential. For its absolute value, we get $a_{Pal\,5} = 2.61^{+0.56}_{-0.44}$\,pc\,Myr$^{-2}$ ($0.81^{+0.17}_{-0.14}\times 10^{-10}$\,m\,s$^{-2}$), which we can convert into a ``circular velocity'' (using Eq.~\ref{eq:vcpal5}) of $V_C(r_{Pal\,5}) = 217^{+22}_{-19}$\,km\,s$^{-1}$, or an equivalent spherically distributed mass yielding $a_{Pal\,5}\,r_{Pal\,5}^2\,G^{-1} = 2.06^{+0.44}_{-0.37}\times10^{11}\msun$. If we assume that Pal\,5 lies in the plane of the Galactic disk at this radius, we arrive at $M(r<r_{Pal\,5})_{disk} = 2.14^{+0.38}_{-0.34}\times10^{11}\msun$. As can be seen in Fig.~\ref{rotationcurve}, our circular velocity estimate bridges a gap between estimates from halo tracers and estimates from disk tracers (Fig.~\ref{rotationcurve}). It fits well into the general trend of the Milky Way rotation curve when combined with the estimates listed below.\\\\
The brightest tracers in the Milky Way halo, and hence the most obvious targets for Milky-Way-mass studies, are the Galactic satellites, i.e.~globular clusters and dwarf galaxies, but also bright stellar tracers for which reasonable distances can be estimated, e.g.~blue horizontal branch stars (BHB). Results from such studies differ widely due to ambiguities in the assumed anisotropy of the tracers. \citet{Wilkinson99} used positions, distances, radial velocities and a few existing proper motions of 27 satellites beyond 20\,kpc to estimate a Milky Way mass of $5.4^{+0.2}_{-3.6}\times10^{11}\msun$ within a radius of 50\,kpc. Similarly, \citet{Watkins10} get masses for the Milky Way inside a radius of 100\,kpc of $M(r<100\,\mbox{kpc})=(0.2-2.3)\times10^{12}\msun$, depending on the assumed orbital anisotropy of the satellites. Their estimate using the anisotropy provided by the (limited) data itself is $(1.4\pm0.5)\times10^{12}\msun$. \citet{Sakamoto03} use satellite galaxies, globular clusters and horizontal branch stars to constrain the mass within the inner 50\,kpc of the Galaxy to be $M(r<50\,\mbox{kpc}) = 5.3^{+0.1}_{-0.4}\times 10^{11}\msun$. \citet{Battaglia05} use similar data to derive an enclosed mass within 120\,kpc of $M(r<120\,\mbox{kpc})=5.4^{+2.0}_{-1.4}\times10^{11}\msun$. \citet{Smith07} use RAVE data to measure a mass for the inner 50\,kpc of $M(r<50\,\mbox{kpc}) = 4.0^{+1.1}_{-0.76}\times10^{11}\msun$ assuming an NFW halo. Analyzing a sample of radial velocity measurements for a large sample of BHB stars in the halo, \citet{Xue08} derive a mass within 60 kpc of $(4.0\pm0.7)\times10^{11}\msun$. \citet{Gnedin10} apply spherical Jeans models to a sample of several hundred BHB and RR Lyrae stars with distances between 25 and 80\,kpc. They determine an enclosed mass of $M(r<80\,\mbox{kpc}) = 0.69^{+0.30}_{-0.12}\times10^{12}\msun$. In a similar way, with dynamical modeling of >4000 BHB stars in the halo with radial velocities from SDSS/SEGUE, but also with allowing for non-sphericity of the tracer distribution function, \citet{Deason12} infer a mass within 50\,kpc of $(4.2\pm0.4)\times10^{11}\msun$, whereas \citet{Kafle12} arrive at a Galactic mass within 25\,kpc of $2.1\times10^{11}\msun$ (without quoted uncertainty). Most recently, \citet{Gibbons14} modeled the Sagittarius stream and arrive at $M(r<100\,\mbox{kpc}) = (4.1\pm0.4)\times10^{11}\msun$, which is at the lower end of the spectrum of mass estimates.\\\\
In Fig.~\ref{rotationcurve}, we also include some mass estimates, or circular velocity estimates, for the Solar environment to get a more complete picture of the Galactic rotation curve. \citet{Koposov10} use the GD-1 stream to constrain the circular velocity at a distance of $8.4$\,kpc to be $(229\pm18)$\,km\,s$^{-1}$. \citet{Bovy12} use APOGEE radial velocity data on disk stars to constrain the local circular velocity at the distance of the Sun from the Galactic center to be $V_C(R = 8.1\,\mbox{kpc}) = (218\pm6)$\,km\,s$^{-1}$. \citet{McMillan11} uses a range of tracers and constraints to arrive at $V_C(R=8.3\,\mbox{kpc}) = 239.2\pm4.8$\,km\,s$^{-1}$. Our estimate for the circular velocity at the Solar radius is $V_C(R_\odot) = 233.0^{+12.7}_{-10.0}$\,km\,s$^{-1}$, and for the Solar radius we get $R_\odot = 8.30^{+0.24}_{-0.25}$\,kpc (see Sec.~\ref{ssec:discussionsolarparameters}).\\\\
Ultimately, tracers in the disk and in the halo have to be combined and modeled altogether. Following such an approach, \citet{Bhattacharjee14} infer a mass within 200\,kpc of $M(r<200\,\mbox{kpc}) = (0.68\pm0.41)\times10^{12}\msun$. They argue that the uncertainties for the Solar galactocentric distance and the circular velocity at the location of the Sun are the two main causes of uncertainties for mass estimates in the inner MW, whereas the missing constraints on the radial anisotropy of the tracers' phase-space distribution in the halo is responsible for the uncertainties at large radii.\\\\
In fact, the mass estimates from the halo tracer studies differ widely, and it has to be assumed that they suffer from the complexity, i.e.~non-sphericity, of the Galactic mass distribution, and of streaming motions among halo stars. From SDSS data, \citet{Bell08} find the stellar halo to be significantly oblate, but furthermore show that a significant fraction of the halo stars (>40\%) are in spatial substructures. This observation is evidence that large part of the stellar halo is formed from infalling satellites (e.g., \citealt{Bullock05}). It is unclear what kind of biases such coherent structures could cause in the modeling of the stellar halo. 

\subsubsection{Shape of the dark halo}
While the coherent debris of Galactic satellites is unpleasant noise for the modeling of the stellar halo, it may offer the most powerful way of measuring the mass distribution of the Milky Way. From our point of view from within the Galaxy, tidal streams like the Sagittarius stream \citep{Ibata94, Dohm01, Majewski03, Belokurov06a, Belokurov14} or the GD-1 stream \citep{Grillmair06b} span huge arcs on the sky, and as coherent phase-space structures simultaneously probe large parts of the Milky Way potential without the need for a tracer density profile or assumptions on orbital anisotropy. Moreover, the hope is that debris streams are also sensitive to the shape of the mass distribution and therefore can give us insights to the 3D mass profile of the Milky Way's dark halo. However, the many studies on the Sagittarius (Sgr) stream have shown that inferring the shape of the dark matter halo from a stream as complex as Sgr is non-trivial.\\\\
Since the first models of the disrupting Sgr satellite by \citet{Johnston95} and \citet{Velazquez95}, the interpretation of its debris has led to many different conclusions. Early numerical models of the Sgr stream were made in spherical halo potentials, and the quantity and quality of data on the discovered debris was not conclusive enough to pinpoint a specific shape of the halo. \citet{Ibata01} pointed out that a large flattening of the halo potential would lead to a significant orbital precession between the leading and the trailing tail, and hence they argued against potential flattenings with $q_z<0.9$. When Sgr was traced along a nearly perfect great circle on the sky in 2MASS data \citep{Majewski03} this hypothesis seemed to be further confirmed. However, \citet{Helmi04a} argued that the Sgr stream was in fact too young to be sensitive the halo shape, and that the halo potential could therefore be either oblate, spherical or prolate. With more kinematic data at hand, \citet{Helmi04b} finally argued for a prolate halo shape, while \citet{Law05} and \citet{Johnston05} argued for an oblate halo to explain the small degree of orbital precession observed in the Sgr debris. With successive detections of new parts of Sgr's tidal debris, this discussion has yet become more complicated. Moreover, \citet{Belokurov06a} and \citet{Koposov12} have found bifurcations in Sgr's leading and trailing stream, which are still not understood. \citet{Fellhauer06} and \citet{Penarrubia10} tried to explain these new features with multiple wraps of the stream around the Milky Way, or with rotation in the Sgr progenitor, respectively. Yet, to the present day there has been no conclusive model that can explain all observed Sgr features.\\\\
Using a large set of collisionless $N$-body simulations, \citet{Law10} reproduced many parts of the observed debris. However, they found the necessary shape of the dark matter halo to be rather peculiar: the observed positions, distances and radial velocities of Sgr debris stars can be explained by a satellite orbiting within a triaxial halo potential that has its minor axis pointing towards the Sun within the disk. That is, the dark matter distribution resembles an oblate halo perpendicular to the Galactic disk, which we see face on when we look towards the Galactic Center. \citet{Deg13} confirmed these findings using also other observational constraints and a different form of parametrization for the Galactic potential.\\\\
However, several problems have been pointed out with such a configuration. \citet{Debattista13} tried to produce a stable galaxy configuration resembling the proposed scenario of \citet{Law10} using a live dark matter halo, and found it to be highly unstable. Moreover, \citet{Pearson14} demonstrated that the thin morphology of the Pal\,5 stream cannot be reproduced in such a triaxial configuration. Such cross checks of stream modeling results are obviously vital; a certain Galactic potential that can produce one specific stream also has to be able to yield models for all other streams. Interestingly, \citet{Pearson14} found that it is easily possible to find a stream model that has the long, thin and curved morphology of the observed Pal\,5 stream in a spherical halo potential. Moreover, the authors point out that a model, which fits the morphology within a spherical halo potential, also naturally reproduces Pal\,5's radial velocity gradient, making a strong case for a nearly spherical dark halo potential. \citet{Ibata13} have in fact shown that it is possible to reproduce the \citet{Law10} observational constraints in a galaxy potential that is spherical. However, they found the necessary potential to be peculiar as well, in that it has to have a rising rotation curve at large radii.\\\\
From cosmological (dark-matter-only) simulations we expect dark matter \textit{density} profiles to be triaxial, where the expected axis ratios lie between 0.4 and 0.6 for the short, intermediate, and long axes \citep{Dubinski91, Allgood06}. While these axis ratios should be prevalent at large galactocentric radii, \citet{Dubinski94} and \citet{Debattista08} showed that condensations of baryons leads to rounder halos in centers of galaxies. \citet{Vera13} therefore suggested a potential form that is triaxial in the outer parts and more spherical in the inner parts. They showed that with such a configuration and an inner potential flattening of about 0.9 it is possible to reproduce a Sgr configuration like the \citet{Law10} model. A flattening/triaxiality that varies with Galactocentric radius may in fact be a way to resolve many conflicts between existing Milky Way mass estimates.\\\\
Yet, Sgr's story is not over: new observational data from \citet{Belokurov14} show that Sgr's tidal debris goes in fact out to 100\,kpc. This debris structure cannot be reproduced by the \citet{Law10} model. While the halo configuration of \citet{Gibbons14} (Milky Way halo with a spherical but sharply cut-off density profile) can reproduce the apocentric distances of observed Sgr debris, and also the precession of Sgr's orbital plane, it has not been shown to reproduce all available observational features. Thus, there is still no model of the Sgr satellite that can reproduce all observational constraints simultaneously.\\\\
The only other stream that has been modeled to infer the shape of the Milky Way's dark halo has used the 63\,deg long GD-1 stream \citep{Koposov10, Bowden15}. The GD-1 stream is very thin but lacks any obvious progenitor satellite, hence it is most probably the stream of a dissolved globular cluster. In contrast to the Sgr satellite it orbits in the inner halo. \citet{Koposov10} estimate the pericenter and apocenter of its orbit to be 14\,kpc and 26\,kpc, respectively. By fitting orbits to sky positions of the debris, distances along the stream, radial velocities, and proper motions, the authors conclude that the best-fitting orbits constrain the dark halo to be nearly spherical. They place a lower limit on the flattening of the dark halo potential, saying that with 90\% confidence the halo flattening is $q_z>0.89$. Recently, \citet{Bowden15} confirmed these results with more sophisticated streakline modeling of the stream.\\\\ 
Our results for the flattening also suggest that the inner 19\,kpc of the Milky Way's dark halo are rather spherical (assuming that our choices for the disk and bulge potentials were reasonable). The results from the different fitting approaches all agree within their uncertainties with each other, and they are actually all compatible with a spherical halo. Our best estimate is $q_z = 0.95^{+0.16}_{-0.12}$, which supports the notion of \citet{Vera13} of having a slightly oblate inner halo, and a possibly less relaxed and probably triaxial halo at larger radii.  On the other hand, our result is also compatible with the nearly spherical ``phantom'' halos predicted by alternative gravity theories \citep{Famaey12, Lughausen14}.\\\\
Yet, the existence of the Sgr dwarf galaxy and the two Magellanic Clouds should remind us that the halo is not necessarily a relaxed structure, but a constantly growing and probably highly substructured potential well. This is also suggested by \citet{Saha09} based on observations of asymmetries in the Galactic \textsc{H\,i} disk. In fact, the assumptions of steadiness and symmetry may be the main reasons for some of the discrepancies pointed out above and may pose the philosophical question of what we actually measure when we model tracers in the Galactic halo. By applying the \texttt{FAST-FORWARD} method to 256 globular cluster streams, which formed in a cosmologically motivated, and time-evolving dark-matter halo of a Milky-Way sized galaxy, \citet{Bonaca14} demonstrated that we can recover the \textit{present-day} mass and shape of the dark matter halo to an accuracy of 5-20\%. This, however, only holds for an ensemble of streams; single streams can show mass biases of up to 50\%. In order to avoid reporting such biased results, detailed comparison to simulated streams on similar orbits in similar potentials, and comparison to results on the same object from other, independent methods should be taken into consideration. Our positive result from the Pal\,5-analog simulation (Sec.~\ref{ssec:resultsanalog}), and the agreement of our fitting results for all other model parameters with estimates from completely independent methods (as shown in the following Sections) give us this necessary confidence in our results out to Pal\,5's apogalactic radius of 19\,kpc.

\subsection{Comparison of Pal\,5 parameters to literature values}

We use five parameters to model Pal\,5 and its orbit. The parameters are its distance to the Sun, its proper motion (2 components), its mass, and its mass loss rate.  

\subsubsection{Heliocentric distance}
Of all the cluster parameters in our modeling, the distance from the Sun to Pal\,5 is the observationally best determined parameter. \citet{Dotter11} use deep HST/WFPC2 data to fit isochrones to Pal\,5's stellar population, and determine a distance of 23.55\,kpc. This estimate comes without an uncertainty, but is supposed to be accurate to within $\pm1$\,kpc (A.~Dotter, private communication). The agreement with our estimate of $23.58^{+0.84}_{-0.72}$\,kpc is remarkable. \citet{Harris96} lists a distance of 23.2\,kpc, which is based on CCD photometry by \citet{Smith86}. This older value is also in agreement with our result.

\subsubsection{Proper motion}
The proper motion of Pal\,5 is relatively unconstrained. \citet{Odenkirchen02} point out two different measurements coming from photographic plates with baselines of four decades each. The estimates for $(\mu_{\alpha}\cos(\delta), \mu_\delta)$ are $(-1.0, -2.7)$\,mas\,yr$^{-1}$ \citep{Scholz98}, and $(-2.55, -1.93)$\,mas\,yr$^{-1}$ (\citealt{Schweitzer93}, revised by K.~Cudworth in 1998). Our best-fit values are $(-2.39^{+0.15}_{-0.17}, -2.36^{+0.14}_{-0.15})$\,mas\,yr$^{-1}$, and lie well within the range of the two measurements. Future proper motion measurements with HST will test this parameter more accurately.\\\\
\citet{Pawlowski12} point out that Pal\,5 and its stream lie within the vast polar structure, which is a coherently rotating disk of satellites and streams that contains most of the Galactic satellites. Our proper motion estimate verifies the notion of \citet{Pawlowski14} that Pal\,5 is in fact counter orbiting within this structure.

\subsubsection{Mass}
Pal\,5's mass is not well determined. The \citet{Harris96} catalogue lists an absolute V-band magnitude of $M_V = -5.17$\,mag, which corresponds to $L/L_\odot = 10^{(M_{V,\odot} -M_V)/2.5}  = 10\,000$. A possible mass-to-light ratio between 0.5 and 3 allows a wide range of cluster masses. Our estimate of $16.0^{+8.5}_{-5.9}\times 10^3\msun$ is well within this range. However, \citet{Odenkirchen02} arrive at an absolute magnitude of only $M_V = -4.77\pm0.20$\,mag, i.e.~a luminosity of $6\,900\,L_\odot$. They also measure the velocity dispersion of Pal\,5 and estimate that their sample is contaminated by a large fraction of binaries (24-63\%). After correcting for this, they arrive at a cluster velocity dispersion between 0.12 and 0.41\,km\,s$^{-1}$, corresponding to a low cluster mass of $5-8\times10^3\msun$, when fit with a dynamical \citet{King63} model. This analysis relies on the assumption that the contribution from the binaries can be clearly separated from the rest of the cluster stars. Uncorrected, the velocity dispersion is $(1.1\pm0.2)$\,km\,s$^{-1}$, in agreement with a recent determination by \citet{Kuzma14} of $(1.3\pm0.4)$\,km\,s$^{-1}$. Using a simple virial argument and Pal\,5's half-light radius of $r_h\approx 20$\,pc, we can estimate the virial mass of Pal\,5 to be $M^{vir}_{Pal\,5} = \kappa\,r_{h}\,\sigma_{3D}^2\,G^{-1} = \kappa\times16\,900\msun$, where $\sigma_{3D} = \sqrt{3}\sigma$ is the three-dimensional velocity dispersion, and $\kappa$ is some proportionality constant, which relates the cluster's half-light radius to the virial radius. For a Plummer model $\kappa \approx 1.3$, which would yield a mass of about $22\,000\msun$. However, this calculation assumes that Pal\,5's density profile is well approximated by a Plummer sphere, that mass follows light in the cluster, and that the cluster is in virial equilibrium. All three of the assumptions may not be fulfilled, since the cluster's surface brightness profile may be better approximated by a low-concentration King model \citep{Odenkirchen02}, the cluster itself is heavily mass segregated and its dark remnant content is unknown \citep{Koch04}, and the cluster may have been tidally shocked recently and brought out of equilibrium \citep{Dehnen04}. The fact that we provide an independent measure on Pal\,5's mass, which does not depend on assumptions about the stellar mass function, dark remnant content, binary fraction and virial state, may help to understand the other mass estimates better. With more kinematic data on the cluster stars and more sophisticated dynamical modeling, our result could be easily tested.

\subsubsection{Mass loss rate}
The mass-loss rate of Pal\,5 has been estimated by \citet{Odenkirchen03} using SDSS star counts and simple assumptions on the drift rate of stars along the tails. Using their lower present-day mass for Pal\,5 of $6\,000\msun$, the authors estimate a rate of 4.9\msun\,Myr$^{-1}$ from the trailing tail, and 3.8\msun\,Myr$^{-1}$ from the leading tail star counts. Our weakly constrained estimate of $7.9^{+5.3}_{-6.4}\msun$\,Myr$^{-1}$ is compatible with these results. When we use their cluster-mass independent estimate of $-\dot{M}/M_{Pal\,5} = (0.7 \pm 0.2)$\,Gyr$^{-1}$ and our present-day mass of $16\,000\msun$, we get $dM/dt = (11.2\pm3.2)\msun$\,Myr$^{-1}$, which is on the upper end of our estimate. Like for the cluster mass, these estimates are completely complementary, as our result does not depend on star counts or assumed drift rates, but is a free parameter in the modeling process. Deeper imaging of the tidal tails will help to assess the mass within the stream and estimate the mass loss rate.

\subsection{Comparison of Solar parameters to literature values}\label{ssec:discussionsolarparameters}

\begin{figure}
\centering
\includegraphics[width=0.45\textwidth]{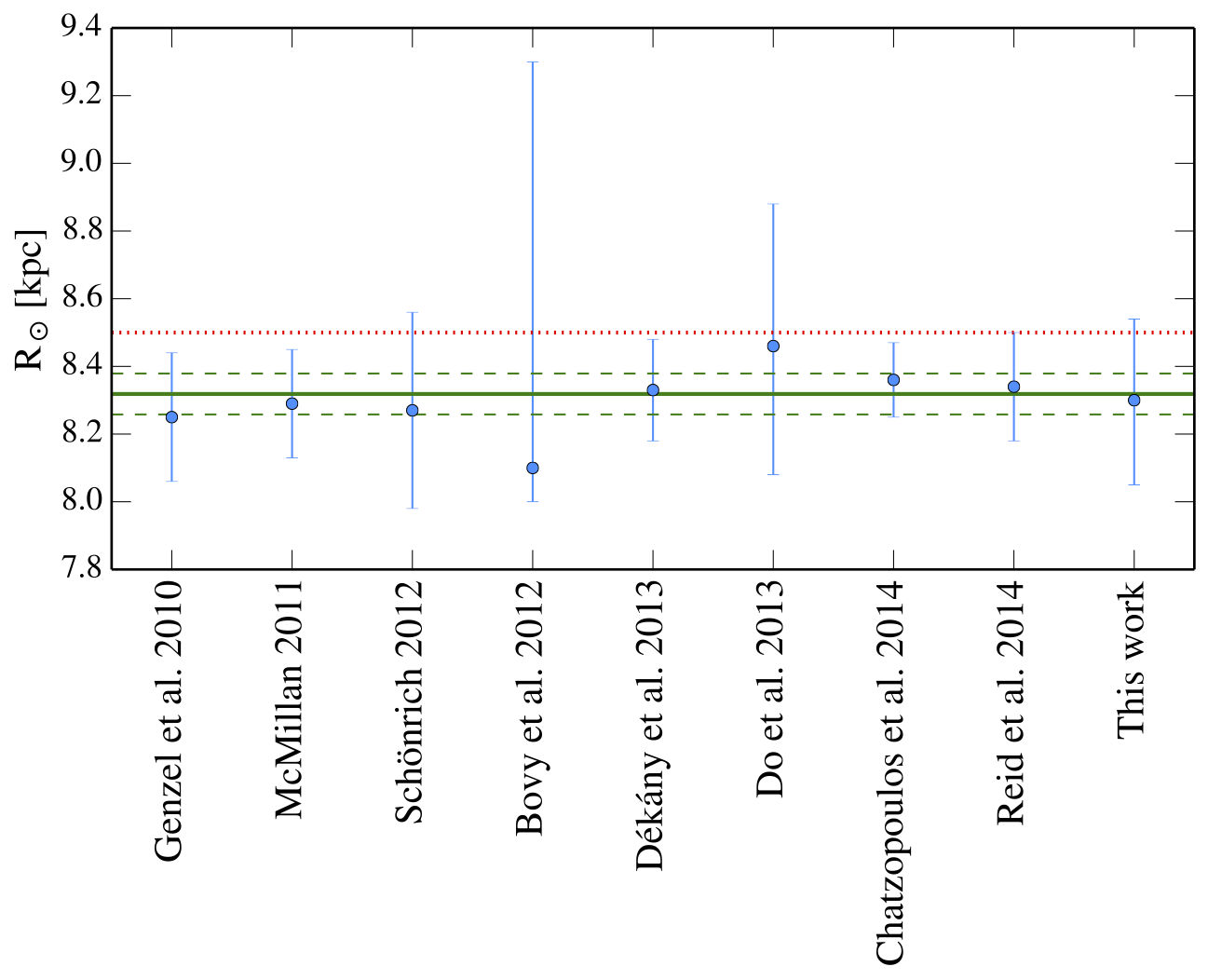}
\caption{The distance of the Sun to the Galactic center, $R_{\odot}$, as estimated by recent studies. Our estimate is given on the far right. It agrees well with the weighted mean of all measurements shown as green solid line with green dashed lines indicating the weighted sample variance. For comparison, the IAU standard of 8.5\,kpc is shown as a red dotted line \citep{Kerr86}.}
\label{distance_GC}
\end{figure}

The two Solar parameters in our modeling are the Galactocentric distance of the Sun, $R_\odot$, and its tangential motion with respect to the Galactic Center, $V_{tan}$. Independent estimates for these quantities exist from various other methods.

\subsubsection{Galactocentric distance}
The distance of the Sun from the Galactic Center has long been unknown to about 10\% uncertainty. With high-precision  measurements of, e.g., the parallax of Sgr A*, the orbital motion of the central S-stars, the dynamics of the nuclear star cluster, or of maser sources in the Milky Way disk, our knowledge on this quantity has increased to better than 3\% uncertainty within recent years (Fig.~\ref{distance_GC}). \citet{Genzel10} summarize the many different direct and indirect estimates for the distance of the Sun to the Galactic Center prior to 2010 to be $R_\odot = (8.25\pm0.19)$\,kpc, where the uncertainty is the variance of the weighted mean of all the measurements. \citet{McMillan11} fit dynamical models to various photometric and kinematical data to arrive at $R_\odot = (8.29 \pm 0.16)$\,kpc. Similarly, just by sophisticated modeling of existing observational data, \citet{Schonrich12} gets $R_\odot = (8.27 \pm 0.29)$\,kpc. More recently, \citet{Dekany13} use distance data on thousands of RR Lyrae stars in the bulge to get $R_\odot = (8.33 \pm 0.05 \pm 0.14)$\,kpc. \citet{Do13} use anisotropic, spherical Jeans models to model 3D kinematic data of the nuclear star cluster, and arrive at $R_\odot = (8.92\pm0.58)$\,kpc, and $R_\odot = 8.46^{+0.42}_{-0.38}$\,kpc when they include data on the orbit of the star S0-2 around Sgr A*. \citet{Chatzopoulos14} improve upon the \citet{Do13} results by using in addition 2\,500 line-of-sight velocities, 10\,000 proper motions, and 200 maser velocities. Moreover they also fit the surface density profile of the nuclear star cluster and allow for rotation. Their modeling yields $R_\odot = (8.30\pm0.09\pm0.1)$\,kpc, and $(8.36\pm0.11)$\,kpc if they include the S-star orbits from \citet{Gillessen09}. Finally, \citet{Reid14} determine $R_\odot = 8.34 \pm0.16$\,kpc  from trigonometric parallaxes of star forming regions in the Galactic disk. All these recent estimates agree very well with our result of $R_\odot = 8.30^{+0.24}_{-0.25}$\,kpc. Moreover, our estimate is in good agreement with the weighted mean of all these measurements listed above, which is $(8.32\pm0.06)$\,kpc (Fig.~\ref{distance_GC}).

\subsubsection{Solar transverse velocity}
\citet{Reid04} determined the tangential motion of the Sun with respect to the Galactic Center from VLBI measurements of the position of Sgr A* to be $\theta_{tan} = 6.379\pm0.026$\,mas\,yr$^{-1}$ in the Galactic plane, corresponding to $30.24\pm0.12$\,km\,s$^{-1}$\,kpc$^{-1}$. Our estimate for this quantity is $\theta_{tan} = 6.43\pm0.46$\,mas\,yr$^{-1}$ ($30.6\pm2.2$\,km\,s$^{-1}$\,kpc$^{-1}$) and hence in good agreement with this strong constraint. Most estimates of the Solar tangential motion, $V_{tan}$, use this measurement of the proper motion of Sgr A* and differ only in the way the distance to the Galactic Center is estimated. \citet{Schonrich12} uses local kinematics of stars from SDSS DR-8 in combination with the Sgr A* measurement to arrive at $V_{tan} = 250\pm9$\,km\,s$^{-1}$. \citet{Bovy12} use APOGEE spectra to model the local Galactic rotation curve. They  estimate $V_{tan} = 242^{+10}_{-3}$\,km\,s$^{-1}$. Most recently, \citet{Reid14} use masers in the Galactic disk to measure $V_{tan} = 252.2\pm4.8$\,km\,s$^{-1}$ independently of the motion of Sgr A*. They also derive a proper motion of Sgr A* from their data, giving $30.57\pm0.43$\,km\,s$^{-1}$\,kpc$^{-1}$. All of these measurements are in agreement with our best-fit value of $V_{tan} = 254\pm16$\,km\,s$^{-1}$. In future modeling of Pal\,5 we can use, e.g., the proper motion of Sgr A* as a strong prior on $\theta_{tan}$, and hence get better constraints on $R_\odot$. For now, it is reassuring to see that we get consistent results with these studies that involve very different data.

\section{Summary and conclusions}\label{sec:conclusions}
We have shown that with sophisticated \textit{streakline} modeling of the tidal tails emanating from the Milky Way globular cluster Palomar\,5, and with very little and publicly-available observational data (SDSS \& radial velocities from the literature), we can tightly constrain 10 different model parameters characterizing the Milky Way's dark matter halo, the cluster Pal\,5 itself, and the Sun's position and velocity within the Galaxy.\\\\
To reproduce the characteristic geometry of the Pal\,5 stream on the sky, we used the \textit{streakline} method outlined in \citet{Kupper12} and \citet{Bonaca14}, which generates restricted three-body models of globular cluster tidal tails in a computationally efficient way and therefore allows for scans of a large parameter space. We compared the \textit{streakline} models to locally overdense regions of color-selected Pal\,5 stars from SDSS DR9 \citep{Ahn12}, which we detected with a \textit{Difference-of-Gaussians} process. The overdensities cover a length of 23\,deg on the sky and show a typical width of 0.1-0.7\,deg. Moreover, we used literature values for the radial velocity of Pal\,5 \citep{Odenkirchen02}, and of 17 radial velocities of giants lying in projection and velocity close to the Pal\,5 stream \citep{Odenkirchen09}. Using a Markov chain Monte Carlo approach, and testing the methodology on a Pal\,5-analog $N$-body stream, we constrain the 10 free model parameters of the mock Pal\,5 with high accuracies and to precisions between 3\% and 20\%. Similar precisions are achieved for the observational data, which are all individually in agreement with independent estimates from completely different and complementary methods (see Sec.~\ref{ssec:discussion}), confirming the validity of our approach.\\\\
Our main findings can be summarized as follows:
\begin{enumerate}
\item Since Pal\,5 orbits within an apogalactic radius of 19\,kpc, our best estimate for the mass of the Milky Way is limited to this radius. We estimate the enclosed mass to be $(2.1\pm0.4)\times10^{11}\msun$, corresponding to an acceleration of $a_{Pal\,5} = 0.81^{+0.17}_{-0.14}\times 10^{-10}$\,m\,s$^{-2}$ at the present-day position of Pal\,5. 
Moreover, we find the halo to have a flattening along the z-axis of $0.95^{+0.16}_{-0.12}$ (Fig.~\ref{obs_tri_halo}).
\item Our models constrain Pal\,5 to have a present-day mass of $M_{Pal\,5} = 16.0^{+8.5}_{-5.9}\times 10^3\msun$, and that it has been losing mass in the past few Gyr with an average mass-loss rate of $dM/dt = 7.9^{+5.3}_{-6.4}\times10^3\msun\,\mbox{Gyr}^{-1}$. Its heliocentric distance is $d_{Pal\,5} = 23.6^{+0.8}_{-0.7}$\,kpc, and the proper motion components we constrain to be $\mu_\alpha\cos(\delta) = -2.39^{+0.15}_{-0.17}$\,mas\,yr$^{-1}$ and $\mu_\delta = -2.36^{+0.14}_{-0.15}$\,mas\,yr$^{-1}$ (Fig.~\ref{obs_tri_cluster}).
\item The transverse velocity of the Sun with respect to the Galactic center is $V_{tan} = 254\pm16$\,km\,s$^{-1}$, and for the Galactocentric distance of the Sun we get $R_\odot = 8.30^{+0.24}_{-0.25}$\,kpc (Fig.~\ref{obs_tri_orbital}). 
\end{enumerate}
From the remarkable precision of these measurements (given the small amount of data), we conclude that long, thin and curved tidal streams like the Pal\,5 stream are exceptionally sensitive to length and velocity scales. The sensitivity to mass and acceleration scales is smaller, but given the high uncertainties of other methods on, e.g., the mass of the Milky Way, it still ranks among the most accurate approaches to mass measurements. More observational data, such as an extended imaging coverage of the Pal\,5 stream beyond the SDSS footprint or proper motions from, e.g., Gaia, as well as the usage of additional prior information on, e.g., the transverse velocity of the Sun from proper motion measurements of Sgr A*, will significantly improve the precision and accuracy that can be achieved. Combining several streams will ultimately unleash the full power of tidal streams in this respect \citep{Deg14}.\\\\ 
By testing different analysis approaches for the calculation of the model likelihoods, we find that radial velocities along the stream are very helpful for constraining model parameters. By adding just 17 velocities, the gain in precision can be as high as 50\% for individual parameters (see Tab.~\ref{tab:accuracies}). We also note that deeper imaging of the Pal\,5 stream would be very helpful, as it would increase the accuracy with which locally overdense regions of the stream can be discovered, and foreground/background fluctuations can be removed. The most pronounced overdensities we currently find lie in the trailing tail at a distance of about 3\,deg from the cluster, and have a significance of $>10\,\sigma$ (see Tab.~\ref{tab:overdensities}), but many of the farther detections are only slightly above the detection limit. Deeper imaging may also help to better characterize the nature of these overdensities, and conclusively differentiate between epicyclic overdensities and, e.g., overdensities produced by past encounters of the Pal\,5 stream with dark matter subhalos \citep{Ibata02, Siegal08, Carlberg09, Yoon11, Carlberg12a, Carlberg12b, Erkal15}.\\\\
We have shown that globular cluster streams are simple and well-understood stellar structures, and that we can model them very accurately. We also demonstrated that globular cluster streams show density variations, which, in parts, can be ascribed to epicyclic motion of tail stars. Thus, cold stellar streams in the Milky Way halo are promising objects, not just for inferring length, velocity and mass scales within the Milky Way, but also for studying the clumpiness of the dark halo, and hence for drawing conclusions on cosmology solely based on simple stellar dynamics.

\section*{Acknowledgements}

The authors would like to thank Sverre Aarseth and Daniel Foreman-Mackey for making their codes publicly available. AHWK thanks Jo Bovy, Rodrigo Ibata, Marla Geha, Adrian Price-Whelan and above all Douglas Heggie for insightful discussions and helpful comments. The authors also thank the rest of the \textit{Stream Team} for valuable feedback, and an anonymous referee for very useful remarks. AHWK would like to acknowledge support through DFG Research Fellowship KU 3109/1-1 and from NASA through Hubble Fellowship grant HST-HF-51323.01-A awarded by the Space Telescope Science Institute, which is operated by the Association of Universities for Research in Astronomy, Inc., for NASA, under contract NAS 5-26555. DWH was partially supported by the NSF (grant IIS-1124794) and the Moore-Sloan Data Science Environment at NYU.\\\\
Funding for SDSS-III has been provided by the Alfred P. Sloan Foundation, the Participating Institutions, the National Science Foundation, and the U.S. Department of Energy Office of Science. The SDSS-III web site is \url{http://www.sdss3.org/}.
SDSS-III is managed by the Astrophysical Research Consortium for the Participating Institutions of the SDSS-III Collaboration including the University of Arizona, the Brazilian Participation Group, Brookhaven National Laboratory, Carnegie Mellon University, University of Florida, the French Participation Group, the German Participation Group, Harvard University, the Instituto de Astrofisica de Canarias, the Michigan State/Notre Dame/JINA Participation Group, Johns Hopkins University, Lawrence Berkeley National Laboratory, Max Planck Institute for Astrophysics, Max Planck Institute for Extraterrestrial Physics, New Mexico State University, New York University, Ohio State University, Pennsylvania State University, University of Portsmouth, Princeton University, the Spanish Participation Group, University of Tokyo, University of Utah, Vanderbilt University, University of Virginia, University of Washington, and Yale University.

\bibliographystyle{apj}
\def\pasa{\ref@jnl{PASA}}               

\end{document}